\documentclass[modern]{aastex62}

\newcommand{\snem}{SN\,2001em}
\newcommand\kms{km\,s$^{-1}$}
\newcommand\nh{cm$^{-2}$}
\newcommand\ergs{erg\,s$^{-1}$}
\newcommand\mdot{$M_\odot$\,yr$^{-1}$}
\newcommand\radio{erg\,cm$^{-2}$\,s$^{-1}$\,Hz$^{-1}$}
\newcommand\chandra{{\em Chandra}}
\newcommand\nustar{{\em NuSTAR}}
\newcommand\xmm{{\em XMM}-Newton}
\newcommand\swift{{\em Swift}-XRT}
\newcommand{\comment}[1]{}

\received{May xx, 2020}
\revised{January xx, 2020}
\accepted{\today}
\submitjournal{ApJ}

\shorttitle{Interaction with a dense shell in SN 2001em}
\shortauthors{Chandra et al.}

\begin{document}

\title{Supernova Interaction with a Dense Detached Shell in  SN 2001em}

\correspondingauthor{Poonam Chandra}
\email{poonam@ncra.tifr.res.in}

\author[0000-0002-0844-6563]{Poonam Chandra}
\altaffiliation{Swarna Jayanti Fellow, Department of Science \& Technology, India}
\affiliation{National Centre for Radio Astrophysics, 
Tata Institute of Fundamental Research\\
Ganeshkhind, 
Pune 411007, India}

\author[0000-0002-9117-7244]{Roger A. Chevalier}
\affiliation{Department of Astronomy, University of Virginia,
P.O. Box 400325\\
Charlottesville, VA 22904-4325, USA}

\author{Nikolai Chugai}
\affiliation{Institute of Astronomy of Russian Academy of Sciences\\
 Pyatnitskaya St. 48,
109017 Moscow, Russia}

\author[0000-0002-0763-3885]{Dan Milisavljevic}
\affiliation{Department of Physics and Astronomy, Purdue University\\
525 Northwestern Avenue,
 West Lafayette, IN 47907}

\author[0000-0001-8532-3594]{Claes Fransson}
\affiliation{Oskar Klein Centre, 
Department of Astronomy, Stockholm University\\
AlbaNova, SE-106 91 Stockholm, Sweden}
\begin{abstract}

We carry out a comprehensive analysis of  supernova  \snem\ covering a period of 19 years since discovery.
 SN 2001em is the oldest supernova known to  have undergone a metamorphosis from a  stripped envelope 
 to an  interacting supernova. 
 An early spectrum indicates it exploded as a Type Ib supernova.
 Later, the ejecta   caught up with a  
 dense circumstellar H-shell, ejected a few thousand years before the explosion, triggering interaction between the supernova ejecta and the dense shell, producing radio, X-ray and H$\alpha$
 emission.
  We use  archival data with the Very Large Array in radio bands and with
 \chandra, \xmm\ and \swift\ in X-ray bands, along with    published H$\alpha$ measurements. We combine these data with our low radio frequency observations with the  Giant Metrewave Radio Telescope  at two epochs covering three frequencies.
 While the observations missed the phase when the shock entered the dense shell, the X-rays indicate that the  shock came out of the dense shell 
 at around 1750 days.  
 The data suggest a forward shock origin of the X-ray  emission.
 Radio data show a spectral inversion at late
 epochs ($> 5000$\,d) at around 3 GHz, which mimics the properties
 of the  central absorbed component seen in SN 1986J.  
 A possible explanation for this component is that the progenitor of \snem\ was a massive binary system which had undergone a period of common envelope evolution.
 The hydrogen envelope from the \snem\ progenitor may have been lost as a result of binary interaction.

\end{abstract}

\keywords{supernovae: general ---  supernovae: individual (SN 2001em)
 ---  circumstellar matter --- techniques: interferometric. --- radiation mechanisms: general --- radiation mechanisms: non-thermal}

\section{Introduction} \label{sec:intro}

Traditional classification of supernovae divides them into  categories, Type II or Type I, based on  presence or absence of hydrogen,
respectively, in their optical spectra \citep{filippenko97}.
A  particular subclass of Type I supernovae,   Type Ib/c supernovae (SNe Ib/c), are believed to  arise from massive 
 stars that have lost their hydrogen envelopes.
While the mechanisms for outer envelope ejections are  poorly understood,  winds \citep{puls+08}, 
episodic ejections \citep{so06, pastorello+07,sq14},   
binary interactions \citep{smith14}, 
and common envelope formation and ejection \citep{cc06,chevalier12} are some of the favoured scenarios.
An extreme example of this subclass of supernovae are Type Ic supernovae, which  not only have shed their hydrogen envelope, 
 but the helium envelope is also lost from the
progenitor. They are also known as stripped envelope supernovae and have also gained attention because of their connection with gamma ray bursts   \citep[GRBs,][]{modjaz+16}. 

In the class of  hydrogen rich Type II  supernovae, the  subclass Type IIn supernovae are identified by their dense   circumstellar interaction, 
and hence are  also called interacting supernovae. 
 SNe IIn  can come from a  variety of progenitors as long as they are surrounded by a dense circumstellar medium (CSM).    
   
  With the advent of more sensitive and automated survey telescopes, a large number of supernovae are being discovered and followed up,
challenging the    traditional picture of supernovae classification. 
The boundaries between various supernova  subclasses are merging. Of particular interest in this work are stripped envelope supernovae (SNe Ib/c) that metamorphose into interacting supernovae (SNe IIn).
The  most straightforward reason
for this metamorphosis is that the supernova ejecta are interacting with the dense hydrogen and/or helium shell(s) that were ejected during
  late stellar evolution. 
 The presence of an early stripped envelope supernova phase implies that the exploding star had a fast wind phase just before the supernova.

The CSM in the vicinity of a supernova is shaped mainly by the winds from the progenitor star before explosion.
Since supernova ejecta velocities are
$10 - 100$ times faster than the CSM wind velocities, one can unravel the pre-explosion progenitor mass-loss history spanning
thousands of years  and possibly see  the signatures of the ejected  hydrogen and/or helium envelope shells during  the late stellar evolution stages.
The age at which the ejecta catch up to the  shells indicates the time before the explosion when the shell was ejected. The ejecta - CSM interaction 
manifests itself mainly in non-thermal radio, thermal/non-thermal X-ray and H$\alpha$ emission,   and
thus provides a unique way to constrain the progenitor evolution by unraveling its mass-loss history. 

\begin{figure*}
\centering
\includegraphics*[angle=0, width=0.95\textwidth]{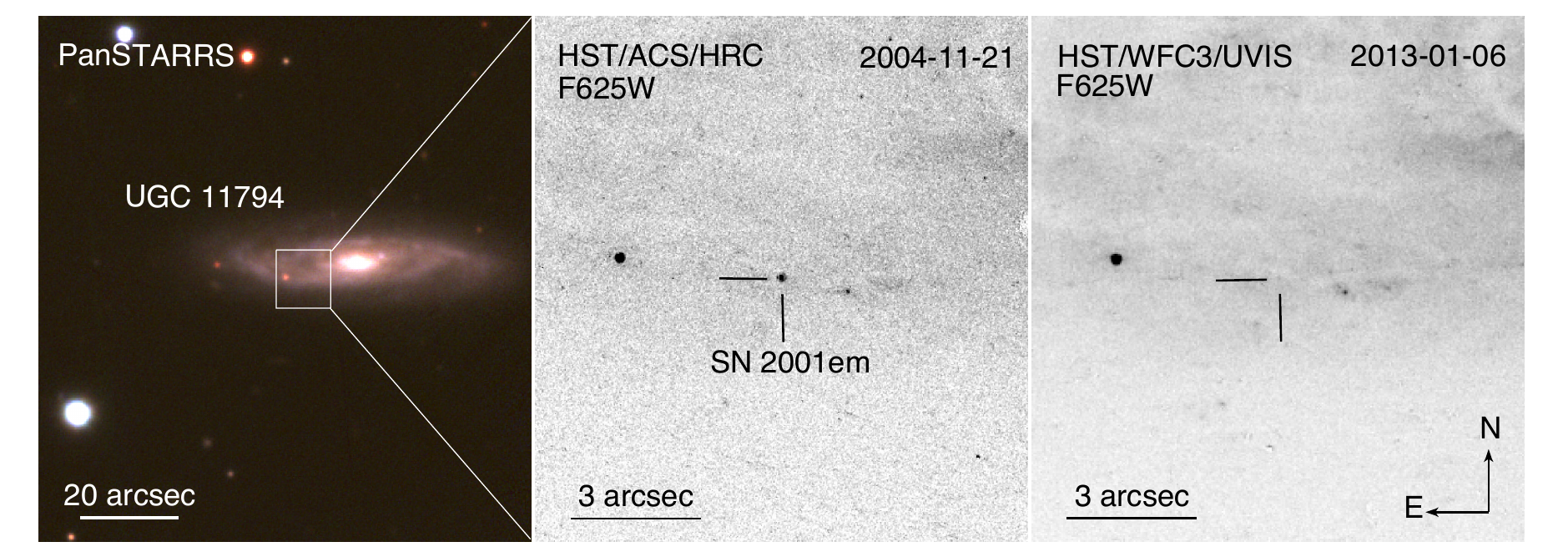}
\caption{Archival optical images of SN\,2001em and its host galaxy UGC\,11797. Left: composite image of the region using PanSTARRS $g$, $r$, and $i$ images. Middle: {\sl Hubble Space Telescope} ({\sl HST}) image (Proposal ID 10272; PI: A.\ Filippenko) obtained with the Advanced Camera for Surveys in the High Resolution Camera mode in the F625W filter sensitive to H$\alpha$, which is the dominant emission line at this epoch. Right: {\sl HST} image (Proposal ID 13029; PI: A.\ Filippenko) obtained with the Wide Field Camera 3 using the same F625W filter, when SN\,2001em is no longer visible.}
\label{fig:findingchart}
\end{figure*}

\snem\ exploded in an outer spiral arm of the galaxy UGC\,11797 \citep[$d=81.2\pm5.7$ Mpc;][Figure~\ref{fig:findingchart}]{grogin+00}.  It was  discovered by \citet{papenkova+01} 
using the Katzman Automatic Imaging Telescope  on 2001 September 20.3
and 21.3  UT.  The estimated explosion date is between September 10.3 UT and September 15.3 UT 
\citep{papenkova+01}; in this paper we adopt 2001 September 13 UT as the explosion date. SN 2001em was classified as a Type Ib/c 
supernova from optical spectra obtained  on 2001 October 20 \citep{fc01}, approximately one month after explosion. 

 \snem\  was   initially a normal  supernova and was  not followed early on with a good cadence. However, due to its classification as a stripped-envelope supernova, some of which are known to be associated with GRBs, 
 \citet{stockdale+04} and 
\citet{pooley+04} carried out  late time radio and X-ray observations to determine whether the supernova 
 harboured a misaligned GRB.  If so, it  could be  detected at late times because the GRB jet would have slowed down with time and have become  quasi-spherical, coming into our line of sight. 
Such a quasi-spherical jet interacting with the surrounding medium is likely to produce detectable radio and X-ray emission. 
\snem\ was indeed detected in both radio and X-ray bands, initially attracting interest in the possibility that a misaligned GRB had  been detected from it \citep{granot+04}.
However, the observed high 8\,GHz radio luminosity ($L_{\rm radio} \sim 2\times10^{28}$\,erg \,s$^{-1}$\,Hz$^{-1}$)
and  the unabsorbed  0.3--10\,keV X-ray luminosity  ($L_{\rm Xray} \sim 10^{41}$\, erg\, s$^{-1}$)  were unprecedented for SNe Ib/c at this age, but were
 comparable to the luminosities of bright SNe IIn
at a similar age \citep{chandra18}.
Later, multiple Very Long Baseline Array (VLBA) measurements  covering epochs between 2004 July to 2004 November    \citep{stockdale+05, paragi+05, bietenholz+05, schinzel+08}  showed no extended
radio emission, although its size  was predicted to be  $\ge2$\,mas in an off-axis GRB model  \citep{granot+04}.  These observations also indicated  the derived velocities
to be  inconsistent with a relativistically expanding radio source in \snem, but were consistent  with an isotropic expansion velocity of $\le 20,000$\,\kms.

The issue was resolved by \citet{cc06}, who  modeled the SN  emission in the framework of the interaction of the supernova ejecta with a dense, $3\,M_\sun$  circumstellar (CS) shell at a distance of $\sim 7\times10^{16}$\,cm formed by vigorous mass loss from the
progenitor star with a rate of $(2-10)\times10^{-3}$\,M$_\sun$\,yr$^{-1}$ at $(1-2)\times10^3$ yr prior to   explosion.
In their model the hydrogen envelope was completely lost and subsequently  swept up and accelerated by the fast wind of the pre-supernova star up to a velocity of $30-50$\,km\,s$^{-1}$.  This model was also supported by the 
optical spectroscopy of the supernova on 2004 May 07 carried out 
by  \citet{soderberg+04} revealing the dominance of
a broad H$\alpha$ emission line with a FWHM velocity of $\sim 1800$ km s$^{-1}$, also later seen by  \citet{vandyk10}  in their 2004 Dec 12 observations with LRIS on the Keck telescope.
The detection of H$\alpha$ emission was not predicted in the off-axis GRB model and was suggestive of a strong interaction with the CSM.
This was further confirmed by \citet{bietenholz+07}, who carried out VLBI observations on
2006 May 27,  5 years after the explosion. They  found that the SN was still unresolved, with a nominal size constraint of  $R=(9\pm15) \times 10^{16}$\,cm, consistent with a non-relativistic expansion of
the radio emitting shock  with an average expansion speed of
$5800\pm10000$  \kms.
The above arguments conclusively established  that the origin of radio, X-ray and H$\alpha$ emission in \snem\ was due to   the  interaction of the ejecta with the dense hydrogen shell.

The results of \citet{cc06}  were based on  early first 1000 days of data and it remained to be seen whether the supernova was still interacting with the dense hydrogen shell or had come out of it.
In this paper, we carry out 
a comprehensive analysis of all available  X-ray 
and radio data, including late time Giant Metrewave Radio Telescope  (GMRT) observations of \snem. 
Although  \snem\ lacks early observations, it is the oldest SN showing a metamorphosis from a stripped-envelope supernova to  an interacting supernova. As more supernovae are being discovered showing such a metamorphosis,
\snem\ presents an opportunity to study the properties of progenitors of such   supernovae for a long time into the pre-explosion. In \S \ref{sec:optical},
we use the available optical data to classify \snem.
The organization of this paper is as follows: the observations are described in \S \ref{sec:observations}. 
We carry out modeling of the data and discuss our main results in \S \ref{sec:results}.
The discussion and  analysis of our results in the context of other similar SNe are in \S \ref{sec:discussion}.   Finally, we give our main conclusions in \S \ref{sec:conclusion}.

\section{Early classification of SN 2001em}
\label{sec:optical}

\begin{figure*}
\centering
\includegraphics*[angle=0, width=0.75\textwidth]{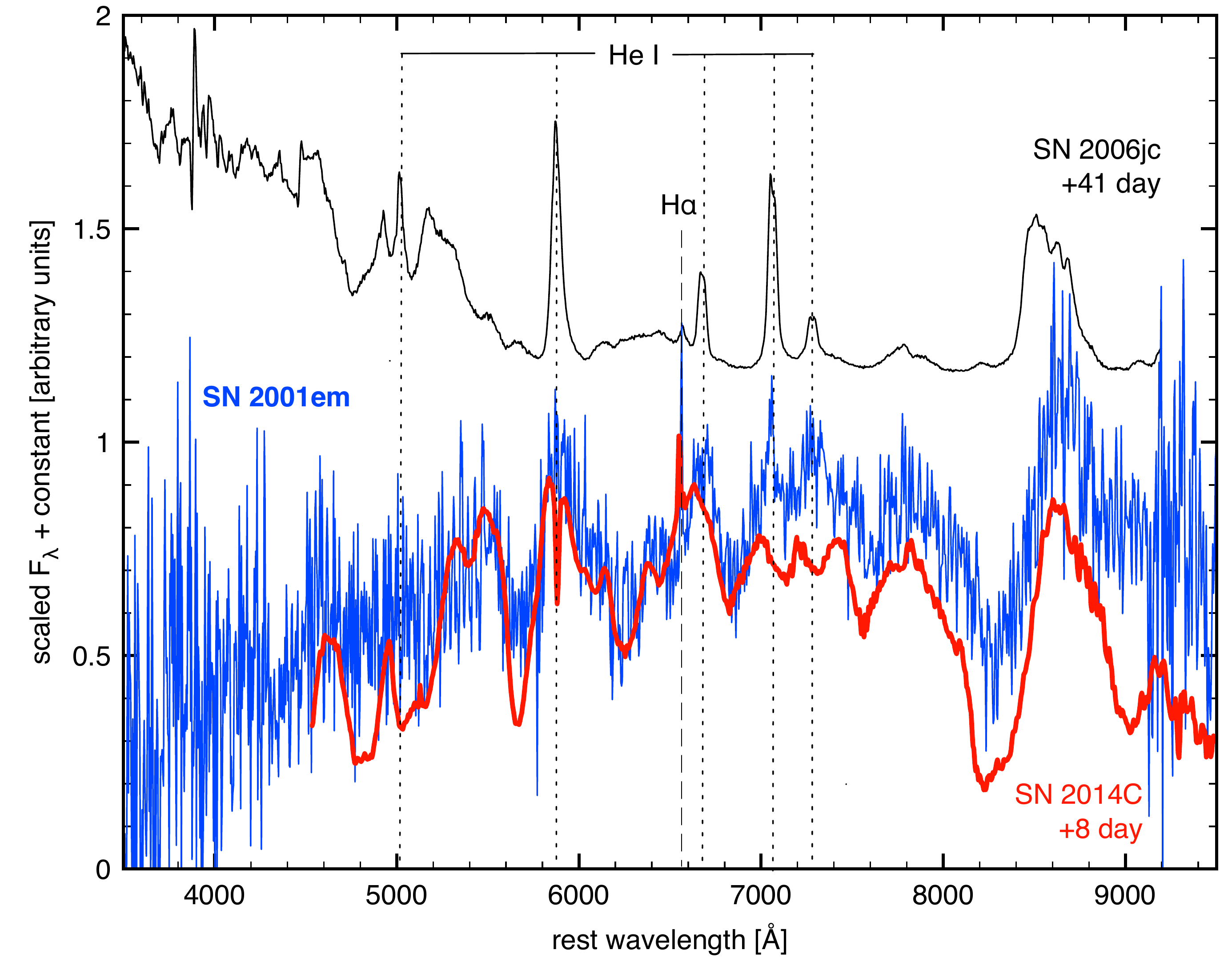}
\caption{The optical spectrum of SN\,2001em shows features consistent with Type Ib supernovae. Excess emission is potentially seen around the locations of \ion{He}{1} lines, especially around the \ion{He}{1} $\lambda$7065 line where it is narrow ($\sim 1000$ km\,s$^{-1}$). This suggests that SN\,2001em may have been interacting with a He-rich CSM environment prior to strong interaction with the H-rich shell. The weak signal-to-noise of the spectrum and lack of additional epochs to follow evolution of the features prevents positive identification.}  
\label{fig:with14c-spectra}
\end{figure*}

There is an ambiguity in the initial classification of \snem\   as either  a Type Ib or a Type Ic supernova. In addition, observations of the intermediate stages between H-deficient to H-dominated emission in \snem\ are limited. We inspected the only known optical spectrum of \snem\ obtained prior to strong interaction, observed by \citet{fc01} and published in \citet{shivvers19}. We confirmed the general Type Ib/c classification but found overall consistency in features more in line with a Type Ib supernova (Figure~\ref{fig:with14c-spectra}). Faithful comparison is complicated by not knowing the exact phase of  \snem\ when the spectrum was obtained, the poor signal-to-noise of the spectrum, and a lack of additional epochs by which to follow the evolution of features. 

Given that the SN\,2001em spectrum was obtained within a month of the estimated explosion date, it is reasonable to assume that the phase was within three weeks of maximum light. We compared the spectrum of SN\,2001em to that of SN\,2014C at the phase +8 days after maximum published in \citet{mauerhan18}.  Most P~Cyg features are shared between the objects, except at wavelengths below 5400~\AA\ where the signal-to-noise is poor and no strong features are observed in SN\,2001em. This further confirms the Type Ib nature of SN\,2001em and suggests a close association between the two supernovae.

However, a noticeable difference between the spectra is that weak excess emission is seen around locations of \ion{He}{1} lines in SN\,2001em that is not 
observed in SN\,2014C. A narrow emission feature centered around 7055~\AA, potentially associated with slightly blueshifted \ion{He}{1} 7065 and having a 
velocity width of  $~\sim 1000$ km\,s$^{-1}$, is particularly conspicuous. Strong narrow helium emission lines are a characteristic 
of Type Ibn supernovae, which are suspected of being explosions that strongly interact with He-rich environments \citep{pastorello07,smith17}. A spectrum of SN\,2006jc is shown \citep{foley07} for 
comparison, with conspicuous \ion{He}{1} lines identified. The normally dominant \ion{He}{1} $\lambda$5876 emission line is noticeably absent from 
\snem. However, for a narrow range of phases,  \ion{He}{1} $\lambda$7065 can be the strongest \ion{He}{1} line in Type Ibn supernovae 
\citep{Karamehmetoglu19}. Although the strong P~Cyg features and lack of blue continuum in SN\,2001em disqualify it as a Type Ibn supernova, the 
consistency in \ion{He}{1} emission locations suggests that \snem\ may have been weakly interacting with He-rich CSM shortly after explosion. If true, this 
scenario would be consistent with the notion that the progenitor star of \snem\ had strong He-rich winds that helped shape the cavity interior to the H-rich 
shell in the millennia approaching core collapse. 

\section{Observations} \label{sec:observations}

\subsection{GMRT radio observations}

We observed   \snem\  with the GMRT on 2004 February 13 at 1.3 GHz. The data were recorded in two polarizations with 16 MHz bandwidth, split into 128  channels.
The observations were carried out in  total intensity mode, and the data were acquired with an integration time of 16.1 sec.
The data were analysed using  the AIPS software.
 Initial flagging and calibration were done using the software FLAGCAL, developed for automatic flagging and calibration for the GMRT data \citep{prasad12}. The flagged and calibrated data were closely inspected;  further flagging and calibration were done manually until the data quality looked satisfactory. The calibration solutions of a single channel were applied to the full bandwidth. Several channels were averaged together to take care of the effect of bandwidth smearing. 
  Fully calibrated data of the target source were imaged using the AIPS task IMAGR. A few rounds of self-calibration were performed.

We also observed  \snem\  with  the upgraded GMRT in frequency band 3 (250--500\,MHz), band  4 (550--900\,MHz) and band 5 (1000--1450\,MHz) in December 2019. The data were recorded with 10\,s integration time and the bandwidth was divided into 2048 channels. The data were analysed using standard Common Astronomy Software Applications (CASA) packages \citep{casa}, following the methodology of \citet{ruta}.

 The flux density of the SN was found by fitting a Gaussian at the SN position using the AIPS task JMFIT.  The details of the GMRT observations and the flux density values can be found in Table \ref{tab:radio}.
 For carrying out the fitting and analysis, we add 10\% uncertainty in quadrature, a typical calibration uncertainty in the
 GMRT data \citep{ck17}.

\subsection{Very Large Array radio observations}

  The Very Large Array (VLA) observations started in October 2003 and continued until July 2020.    The data from the years 2003  to 2008 were obtained with the old VLA covering the frequency range $1.4-22.5$\,GHz. 
 The data were taken in standard continuum mode
with  a total bandwidth  $2 \times 50$ MHz.  We used the  
Astrophysical Image Processing System (AIPS) 
to analyse the VLA data. We analysed  all    publicly available archival VLA data in order to obtain  consistent flux density values.
 Our reanalysed values match with the published flux densities whenever available.

The later data were taken with the upgraded Karl G. Jansky Very Large Array (JVLA). We carried out JVLA data analysis  using the standard packages within the Common Astronomy Software Applications package (CASA).
The details of observations and the flux density values can be found in Table \ref{tab:radio}.
We add 10\% error in the quadrature to the flux density values for analysis purposes, a typical uncertainty in the flux density calibration scale at the observed frequencies
\footnote{https://science.nrao.edu/facilities/vla/docs/manuals/oss/performance/fdscale}.

\begin{longrotatetable}
\begin{deluxetable*}{lccccRRl}
\centering
\tablecaption{Radio observations of SN 2001em} \label{tab:radio}
\tablecolumns{8}
\tablewidth{0pt}
\tablehead{
\colhead{Date of } &  \colhead{Telescope} &  \colhead{Proposal} &  \colhead{Age\tablenotemark{a}} &  \colhead{Frequency} & \colhead{Resolution} & \colhead{Flux density} & \colhead{Ref.}  \\
\colhead{Observation (UT)}   &
 \colhead{Configuration\tablenotemark{b}} & \colhead{ID} & \colhead{(day)} & \colhead{(GHz)} & \colhead{$''\times''$} & \colhead{mJy} & \colhead{}
}
\startdata
2003 Oct 17  & VLA:B & AS779 & 764 &  8.46 & 0.89''\times0.62''& 0.899\pm0.036 & This paper\\
2004 Jan 31 & VLA:BC & AW617 & 870 & 1.43 & 16.72''\times 4.80'' & <0.654 & This paper\\
'' & '' & ''& ''&4.86 & 4.49''\times 1.40'' & 1.596\pm0.081 & This paper\\
'' & ''& ''& ''&  8.46  & 2.25''\times 2.12'' & 1.340\pm0.043 & This paper\\ 
'' & ''& ''& ''&       14.94    & 1.63''\times 0.43'' & 0.859\pm0.219 & This paper\\
2004 Feb 13 & GMRT & 05PCA02 & 883 & 1.29 & 3.92''\times2.39'' & <0.478& This paper\\
2004 Mar 23 & VLA:C& AW617&  922 & 1.43   & 17.77''\times12.99'' & <1.154 & This paper\\
'' & ''& ''& '' &      4.86  & 4.13''\times3.80'' & 1.552\pm0.043 & This paper\\
''& ''& ''& '' &  8.46    & 2.58''\times 2.12'' & 1.595\pm0.042 & This paper\\
''& ''& ''& '' &  14.94  & 1.49''\times 1.23'' & 0.958\pm 0.140 & This paper\\
2004 Apr 20 & VLA:C  & AW624 & 950 &  1.43 & 14.45''\times 13.72'' & <0.645 & This paper\\
 ''& ''& ''& '' &  4.86  &  4.18''\times 3.90'' &  1.784\pm0.054  & This paper\\
''& ''& ''& '' &  8.46  &  2.39''\times 2.30'' &  1.655\pm0.054  & This paper\\
 2004 Jul 01 & VLBA & $\ldots$  & 1022 & 8.46 & 0.0019''\times0.0008'' & 1.8\pm0.2 & Stockdale$+04$\\
2004 Sep 10  & VLA:A & AW642 & 1093 &  1.43  &  1.30''\times 1.26 ''& 0.416 \pm0.064 & This paper\\
 ''& ''& ''& '' &  4.86  &  0.38''\times 0.37'' &  1.773\pm 0.062 & This paper\\
 ''& ''& ''& '' &  8.46  &  0.21''\times 0.21'' &  1.678\pm 0.047 & This paper\\
 ''& ''& ''& '' &  14.94  &  0.13''\times 0.12'' &  0.826\pm 0.205 & This paper\\
 ''& ''& ''& '' &  22.46  &  0.09''\times 0.08'' &  0.897\pm0.118  & This paper\\
 2004 Nov 22 & VLBA & $\ldots$  & 1166 & 8.46 & 0.0009''\times0.0009'' & 1.5\pm0.1 & Bietenholz$+04$\\
  2005 Feb 07& VLA:AB& AW641& 1243&  1.43  &  4.42''\times 1.42'' &  0.862\pm0.169  & This paper\\
 ''& ''& ''& '' &  4.86  &  1.31''\times 0.45'' &  1.752\pm0.267  & This paper\\
 ''& ''& ''& '' &  8.46  &  0.76''\times 0.25'' &  1.430\pm0.060  & This paper\\
 2005 Feb 25  & VLA:B & GB053 & 1261&  1.43  &  4.74''\times 4.49'' & 0.845 \pm0.267 & This paper\\
 ''& ''& ''& '' &  8.46  &  0.78''\times 0.73'' &  1.456\pm 0.085 & This paper\\
 ''& ''& ''& '' &  14.94  & 0.45''\times 0.40'' &  <0.990 & This paper\\
 ''& ''& ''& '' &  22.46  &  0.31''\times 0.26'' &  1.282\pm0.288  & This paper\\
2005 Mar 11 & e-VLBI  &  $\ldots$ & 1275  &  1.6  & 0.0024''\times 0.0024'' &  \sim 0.88  & Paragi$+05$\\
2005 Mar 24 & VLA:B  &  AW641 & 1288  &  1.43  &  4.30''\times 4.09'' &  1.037\pm0.235  & This paper\\
 ''& ''& ''& '' &  4.86  &  1.26''\times 1.20'' &  2.161\pm0.074  & This paper\\
 ''& ''& ''& '' &  8.46  &  0.77''\times 0.64'' &  1.601\pm0.051  & This paper\\
 2005 Apr 26 & VLA:B  &  AW641 & 1321  &  1.43  &  4.38''\times 3.96'' &  1.039\pm0.146  & This paper\\
 ''& ''& ''& '' &  4.86  &  1.26''\times 1.18'' &  2.001\pm0.070 & This paper\\
 ''& ''& ''& '' &  8.46  &  0.72'\times 0.68'' &  1.477\pm0.037  & This paper\\
2005 May 23 & VLA:B  &  AW641 & 1348  &  1.43  &  4.26''\times 3.85'' &  1.307\pm0.304  & This paper\\
 ''& ''& ''& '' &  4.86  &  1.20''\times 1.13'' &  1.694\pm0.138  & This paper\\
 ''& ''& ''& '' &  8.46  &  0.67''\times 0.65'' &  1.349\pm0.082  & This paper\\
2005 Jun 13 & VLA:BC & AW642 & 1369 & 14.94 & 1.47''\times0.44'' & 1.274\pm0.164 & This paper\\
''& ''& ''& ''& 22.46 & 0.95''\times0.29'' & 0.995\pm0.101 & This paper\\
2005 Jun 14  & VLA:BC & AW647 & 1370&  1.43  &  14.80''\times 4.87'' & 1.352 \pm0.119 & This paper\\
 ''& ''& ''& '' &  4.86  &  4.20''\times 1.43'' &  1.968\pm 0.049 & This paper\\
 ''& ''& ''& '' &  8.46  &  2.64''\times 0.78'' &  1.523\pm 0.033 & This paper\\
 ''& ''& ''& '' &  22.46  &  1.00''\times 0.30'' &  0.788\pm0.110  & This paper\\
2005 Jul 12 & VLA:C  &  AW641 & 1398  &  1.43  & 14.52''\times 13.85'' &  1.522\pm0.284  & This paper\\
 ''& ''& ''& '' &  4.86  &  4.08''\times 3.86'' &  1.931\pm0.059  & This paper\\
 ''& ''& ''& '' &  8.46  &  2.39''\times 2.26'' &  1.432\pm0.054  & This paper\\
2005 Aug 12 & VLA:C  &  AW641 & 1429  &  4.86  &  4.00''\times 3.70'' &  1.883\pm0.039  & This paper\\
 ''& ''& ''& '' &  8.46  &  2.34''\times 2.21'' &  1.443\pm0.033  & This paper\\
2006 Apr 17  & VLA:A & AW675 & 1677&  1.43  &  1.36''\times 1.12'' & 1.214 \pm0.105 & This paper\\
 ''& ''& ''& '' &  4.86  &  0.49''\times 0.33'' &  1.586\pm 0.073 & This paper\\
 ''& ''& ''& '' &  8.46  &  0.23''\times 0.19'' &  1.016\pm 0.058 & This paper\\
 ''& ''& ''& '' &  22.46  &  0.09''\times 0.07'' &  0.649\pm0.127  & This paper\\
 2006 May 27 & VLBI & $\ldots$ & 1717 & 8.4 & 0.0017"\times0.0008" & 1.05\pm0.06 & BB$+07$\\
2006 May 27 & VLA:AB & BB219 & 1717 & 1.43 & 12.70''\times2.16'' & 1.282\pm0.205 & This paper\\
''& ''& ''& ''& 22.46 & 0.76''\times0.12'' & <0.765 & This paper\\
2006 Aug 30& VLA:B & AW679 & 1812 & 1.43  & 4.39''\times4.04'' & 1.606\pm0.081 & This paper\\
 ''& ''& ''& '' &  4.86  &  1.30''\times 1.20'' &  1.063\pm 0.047 & This paper\\
 ''& ''& ''& '' &  8.46  &  0.71''\times 0.69'' &  0.749\pm 0.038 & This paper\\
2007 Feb 04 & VLA:CD & BG162 & 1970& 8.46 & 6.82''\times3.15'' & 0.715\pm 0.173 & This paper\\
2007 Feb 18 & VLA:CD & STUDEN & 1984 & 1.43 & 42.31''\times37.88'' & <1.337 & This paper\\
"& "& "& ''& 4.86 & 11.74''\times10.31'' & 1.019\pm0.165 & This paper\\
''& ''& ''& '' & 22.46 & 2.54''\times2.18'' & 0.453\pm0.083 & This paper\\
2007 May 29 & VLA:A & AW709 & 2084 & 1.43 &  1.47''\times1.28'' & 1.075\pm0.050 & This paper\\
 ''& ''& ''& '' &  4.86  &  0.42''\times 0.34'' &  0.913\pm 0.095 & This paper\\
  ''& ''& ''& '' &  14.94  &  0.13''\times 0.12'' &  <0.780 & This paper\\
2008 Sep 25 & VLA:A & AS961 & 2569 & 4.86 & 0.41''\times0.37'' & 0.604\pm0.048 & This paper\\
2008 Oct 01 & VLA:A & AS961 & 2575 & 1.43 & 1.24''\times1.22'' & 0.939\pm0.095 & This paper\\
'' & '' & '' & '' & 8.46 & 0.22''\times0.21'' & 0.298\pm0.030 & This paper\\
'' & '' & '' & '' & 22.46 & 0.10''\times0.08'' & 0.336\pm0.099 & This paper\\
2016 Aug 26 &  VLA:B  & SH746 & 5461 & 3.04 & 2.01''\times1.71'' & 0.062\pm0.016 & This paper\\
''  & ''  & ''  & ''  & 8.55 & 0.74''\times0.63'' & 0.095\pm0.012 & This paper\\
''  & ''  & ''  & ''  & 9.04 & 0.71''\times0.59'' & 0.095\pm0.009 & This paper\\
''  & ''  & ''  & ''  & 11.06 & 0.60''\times0.49'' & 0.104\pm0.011 & This paper\\
2019 Dec 22 & uGMRT & 37\_097 & 6674 &  0.40 & 8.22"\times5.44" & <0.204 & This paper\\
''  & ''  & ''  & ''  & 0.65 & 4.24"\times3.62"& 0.149\pm0.051 & This paper\\
2019 Dec 23 & ''  & ''  & 6675 & 1.26  &2.57"\times1.66" & 0.107\pm0.045 & This paper\\
2020 Jun 29 & JVLA & 20A\_124 & 6865 &  6.0 & 2.02"\times0.97" & 0.050\pm0.010 & This paper\\
2020 Jun 29 & JVLA & 20A\_124 & 6865 &  10.0 & 0.64"\times0.59" & 0.065\pm0.010 & This paper\\
2020 Jun 30 & JVLA & 20A\_124 & 6866 &  32.99 &  0.19"\times 0.17" & 0.056\pm0.011 & This paper\\
2020 Jul 05 & JVLA & 20A\_124 & 6871 & 15.30 & 0.44"\times0.36" & 0.075\pm0.008 & This paper\\
2020 Jul 15 & JVLA & 20A\_124 & 6881 &  2.81 & 2.13"\times2.02" & <0.075 & This paper\\
\enddata
\tablenotetext{a}{The age is calculated using 2001 September 13 (UT) as the date of explosion.}
\tablenotetext{b}{Array configuration relevant only for VLA.}
\tablecomments{Stockdale$+04\equiv$ \citep{stockdale+04}, Bietenholz$+04\equiv$ \citep{bietenholz+04},  Paragi$+05\equiv$ 
\citep{paragi+05},BB$+07\equiv$ \citep{bietenholz+07}}
\end{deluxetable*}
\end{longrotatetable}

\subsection{{\it Chandra} X-ray observations}

We obtained  {\it Chandra} archival data on \snem\ at 5 epochs between 2004 April and 2016 August. 
We extracted  the {\it Chandra} spectra, response and ancillary matrices using 
Chandra Interactive Analysis of Observations software \citep[CIAO;][]{fruscione+06}, using task
\texttt{specextractor}. The 
CIAO version 4.6 along with CALDB version 4.5.9 was used for this
purpose.  The  HEAsoft\footnote{\url{http://heasarc.gsfc.nasa.gov/docs/software/lheasoft/}}
package Xspec version 12.1 \citep{arnaud+96} was used to carry out the spectral analysis. 

The spectra were fitted with a thermal bremsstrahlung model and the temperature was fixed at 20 keV.  Due to low counts, only 5 channels were averaged and 
we used  maximum likelihood statistics for a Poisson distribution, i.e.  
the $c$-statistics\footnote{The $c$-statistics  minimizes the C value defined as $C=-2\Sigma_i(M_i-D_i+D_i(\ln D_i-\ln M_i))$,
where $M_i$ and $D_i$ are defined as  the model-predicted
and observed counts in each spectral bin $i$, respectively. Errors in the parameters in $c$-statistics
are estimated as
in the $\chi^2$ method, i.e. $\Delta C=C -C_{\rm min}$, where $C_{\rm min}$ is the minimum best-fit $C$ 
value obtained.}  \citep{cash79}.
 At the last  epoch, we do not detect X-ray emission from the supernova and 
the upper limit was obtained using a thermal bremsstrahlung model with temperature 20 keV and a column density of $1.5\times10^{21}$ cm$^{-2}$.
Table \ref{tab:xrayobs} lists various model fit values and fluxes of the SN at various epochs.

\subsection{{\it XMM-Newton} X-ray observations}

{\it XMM-Newton} observed \snem\ on 2006 June for an exposure of around 25 ks. 
To extract the spectra and response matrices for the  {\it XMM-Newton}  data,
the Scientific Analysis System (SAS) version 
12.0.1 and its standard commands were used. 
The software 
\texttt{Xspec} 
was used to carry out the spectral analysis, in a similar way to the {\it Chandra} analysis. 
The details are listed in Table \ref{tab:xrayobs}.

\subsection{{\it Swift}-XRT observations}

The X-ray telescope (XRT) onboard {\it Swift} observed \snem\ at various epochs during 2008 January to 2016 January. 
The \texttt{xselect} program of 
HEAsoft 
package  was used to create the spectra and images. Observations closely spaced in time were combined. 
None of the {\it Swift}-XRT observations resulted in detection. The 0.3--10.0\,keV flux  was obtained from  the $3\sigma$ upper limits on count rates by assuming a thermal plasma of 20 keV
and a column density of $1.5\times10^{21}$\,\nh.

In Table \ref{tab:xrayobs}, we list all X-ray observations.

\begin{longrotatetable}
\begin{deluxetable*}{lccccrcr}
\tablecaption{X-ray observations of SN 2001em  \label{tab:xrayobs}}
\tablecolumns{8}
\tablewidth{0pt}
\tabletypesize{\scriptsize}
\tablehead{
\colhead{Date of } &  \colhead{Telescope} &  \colhead{Proposal} &   \colhead{Exposure} &  \colhead{Age\tablenotemark{a}} & \colhead{Count rate} & \colhead{Column density} & \colhead{Unabs. Flux\tablenotemark{b}}  \\
\colhead{Observation (UT)}   &
 \colhead{} & \colhead{ID} & \colhead{(ks)} & \colhead{(day)} &   \colhead{(cts s$^{-1}$)} & \colhead{ (cm$^{-2}$)} & \colhead{(erg s$^{-1}$ cm$^{-2}$)} 
}
\startdata
2004 Apr 04 & {\it Chandra} & 5306 & 14.57 & 934 & $(10.55\pm0.88)\times10^{-3}$ & $(2.69\pm0.59)\times10^{21}$ & $(1.58\pm0.13)\times10^{-13}$ \\
2006 Jun 14 & {\it XMM-Newton} &  0405730101 & 19.70 & 1735 & $(38.49\pm5.04)\times10^{-3}$ & $(1.35\pm0.73)\times10^{21}$ & $(1.56\pm0.23)\times10^{-13}$\\
2007 Apr 23 & {\it Chandra} & 7606 & 4.98 & 2048 & $(5.29\pm1.10)\times10^{-3}$ & $(2.08\pm1.51)\times10^{21}$ & $(7.77\pm1.69)\times10^{-14}$ \\
2008 Jan 03 &  {\it Chandra} & 9094 & 4.98 & 2303 & $(3.65\pm0.90)\times10^{-3}$ & $(1.71\pm1.78)\times10^{21}$ & $(4.84\pm1.37)\times10^{-14}$ \\
2008 Jul 20 & {\it Swift}-XRT & 37982001 & 2.20 &  2502 & $<5.06\times10^{-3}$ & $1.7\times10^{21}$ (fixed) &  $<3.05\times10^{-13}$\\
2009 Aug 22   & {\it Chandra} & 11227  & 10.05 & 2900 & $(2.96\pm0.61)\times10^{-3}$ & $(1.52\pm1.29)\times10^{21}$ & $(3.67\pm0.81)\times10^{-14}$ \\
2012 Jan 11 & {\it Swift}-XRT & 45809001 & 0.87 &  3772 & $<15.9\times10^{-3}$ & $1.5\times10^{21}$ (fixed) &  $<9.62\times10^{-13}$\\
2012 Jun 17 &  {\it Swift}-XRT & 45809002 &1.26 &  3930 & $<10.8\times10^{-3}$ & $1.5\times10^{21}$ (fixed) &  $<6.53\times10^{-13}$\\
2012 Oct 03--Oct 13 & {\it Swift}-XRT & 45809003-004 &2.16 &  $4043\pm5$ & $<7.16\times10^{-3}$ & $1.5\times10^{21}$ (fixed) &  $<4.33\times10^{-13}$\\
2013 Jan 08--Jan 10 & {\it Swift}-XRT & 45809005-007 &3.99 &  $4137\pm1$ & $<3.99\times10^{-3}$ & $1.5\times10^{21}$ (fixed) &  $<2.05\times10^{-13}$\\
2013 Apr 04  & {\it Swift}-XRT & 45809008 & 0.77 &  $4221$ & $<15.1\times10^{-3}$ & $1.5\times10^{21}$ (fixed) &  $<9.13\times10^{-13}$\\
2013 Jun 25 & {\it Swift}-XRT & 45809009 &4.01 &  4303 & $<5.46\times10^{-3}$ & $1.5\times10^{21}$ (fixed) &  $<3.30\times10^{-13}$\\
2015 Jan 10 & {\it Swift}-XRT & 45809010 &  0.46 &  4867 & $<24.9\times10^{-3}$ & $1.5\times10^{21}$ (fixed) &  $<1.51\times10^{-12}$\\
2015 Apr 08--Apr 22 & {\it Swift}-XRT & 45809011-013 &1.04 &  $4962\pm7$ & $<11.2\times10^{-3}$ & $1.5\times10^{21}$ (fixed) &  $<6.77\times10^{-13}$\\
2015 Jun 29--Jul 04 & {\it Swift}-XRT & 45809014-015 &0.47 &  $5040\pm2$ & $<33.1\times10^{-3}$ & $1.5\times10^{21}$ (fixed) &  $<2.00\times10^{-12}$\\
2016 Jan 06 & {\it Swift}-XRT & 45809010 &  0.20 &  5228 & $<58.5\times10^{-3}$ & $1.5\times10^{21}$ (fixed) &  $<3.54\times10^{-12}$\\
2016 Aug 19 &  {\it Chandra} & 18031 & 14.86 & 5454 & $<8.10\times10^{-4}$ & $1.5\times10^{21}$ (fixed) &  $<1.45\times10^{-14}$\\
\enddata
\tablenotetext{a}{The age is calculated using 2001 September 13 (UT) as the date of explosion.}
\tablenotetext{b}{Unabsorbed flux in 0.3--10~keV energy range.}
\tablecomments{We have fixed the temperature to 20 keV for all observations to derive the SN 2001em flux. }
\end{deluxetable*}
\end{longrotatetable}

\section{Modeling and results} \label{sec:results}

\subsection{Visual inspection of radio data}
\label{sec:visual}

\begin{figure*}
\centering
\includegraphics*[angle=0, width=0.79\textwidth]{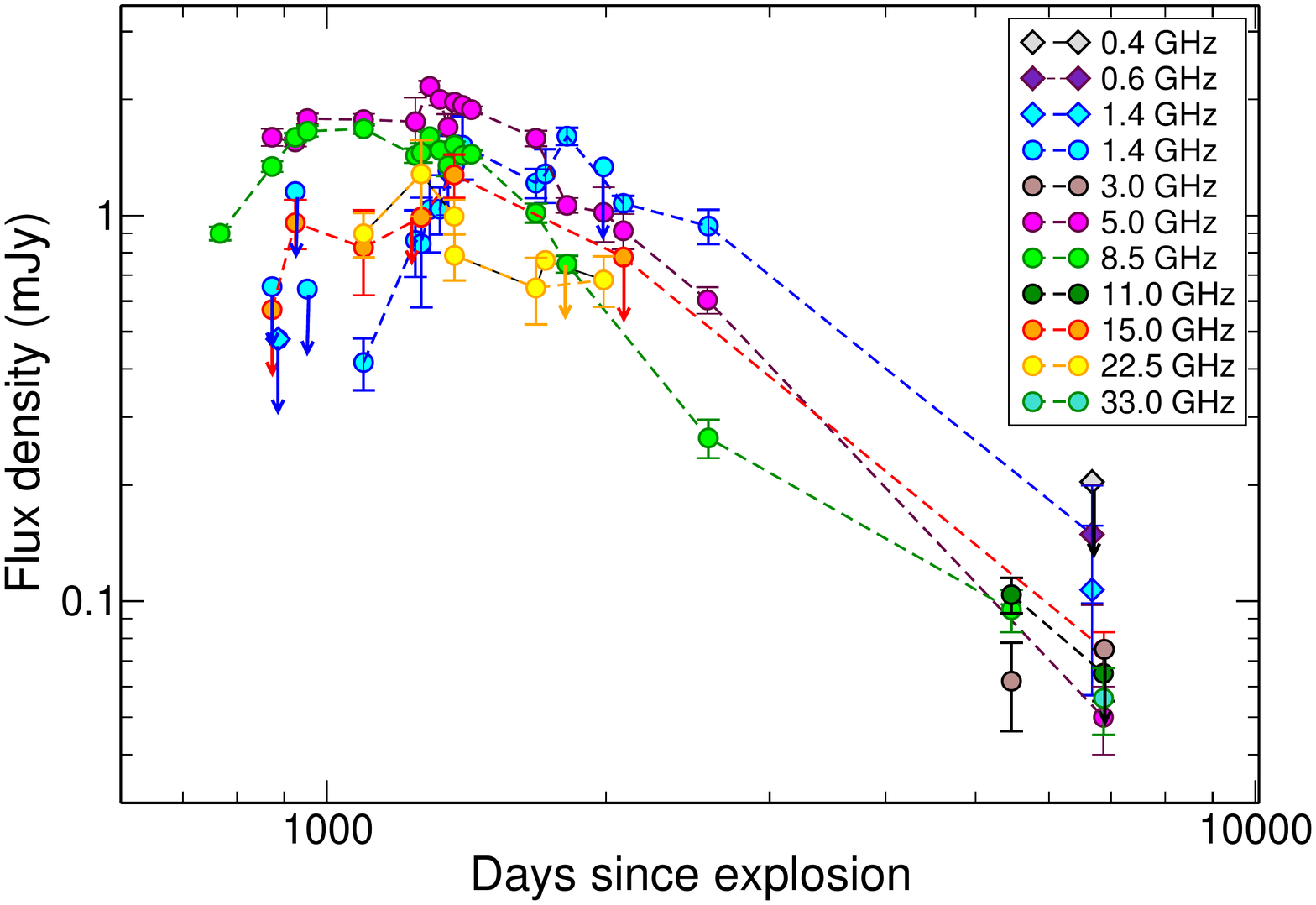}
\includegraphics*[angle=0, width=0.79\textwidth]{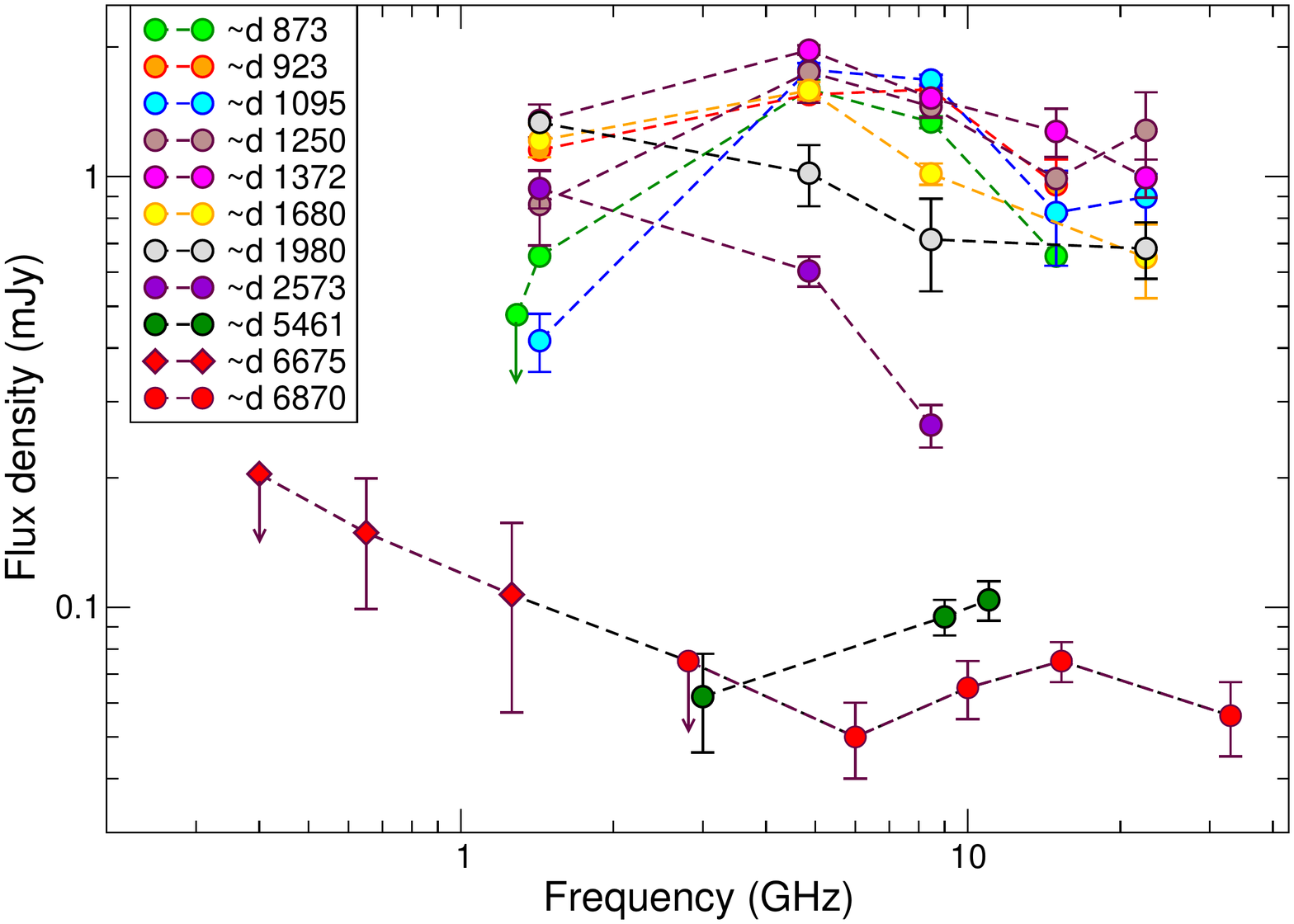}
\caption{{\it Top panel:} The radio light curves of \snem\ at various frequencies. 
{\it Bottom panel:} Radio spectra of \snem\  at various epochs.  The last epoch represents spectra at $\sim 6000\pm500$ days. In both the plots, the circles represent VLA data and the diamonds represent GMRT data. 
\label{fig:radiodata}}
\end{figure*}

 In Fig. \ref{fig:radiodata}, we plot \snem\ light-curves covering frequencies between  0.4\,GHz to 22.5\,GHz and near-simultaneous spectra between epochs 
 873 days to  $\sim$ 6870 days.
The light curves make a transition from  optically thick to optically  thin phase  at all observed frequencies, with 
peak of the light curve  progressively moving to later times at lower frequencies.
At  1.4 GHz, \snem\  turned on at around 1100 days and peaked at $\sim 1700$ days. 
The spectra  seem to follow the standard expected evolution until day 2573, as the peak of the spectra move to progressively lower frequencies at later epochs and eventually 
evolve to the optically thin phase  at all observed frequencies, indicating the absorption peak has moved below $1.4$\,GHz. The peak spectral luminosities at different radio frequencies lie within $(1.1-1.8 ) \times10^{28}$\,\radio. This places \snem\
 among the most  luminous radio supernovae \citep{chandra18}.

 The most unexpected features are in the spectra post 5000 days.  
Since the spectra are at  late epochs and the  fractional change in the supernova age is only 10\% between these two epochs, we combine the 
Dec 2019 and mid 2020 data.  
The spectrum declines at low frequencies but 
starts to rise again at  higher frequencies, and then making a second turn at around 15 GHz (Fig. \ref{fig:radiodata}).
   We discuss this feature in the next section (\S \ref{sec:discussion}). 

\subsection{Radio emission models}
\label{sec:radiomodel}

Radio emission mainly arises from the forward shock and  is synchrotron in nature. We use the self-similar 
synchrotron emission models developed  by \citet{chevalier82a, cf17}, where the shock radius evolves as power-law in time, i.e.
 $r\propto t^m$, $m$ is the 
 shock deceleration parameter  and is connected to the outer ejecta density profile  index $n$ (in $\rho_{\rm ej} \propto r^{-n}$) as
$m=(n-3)/(n-s)$. Here $s$ is  the index of the power-law density profile  in the unshocked CSM, $\rho_{w} $, created by the stellar wind of the progenitor star ($\rho_{w} \propto r^{-s}$).  
A steady stellar wind   is generally assumed for the progenitor wind, i.e. $s=2$. However, SNe IIn are known to deviate significantly from the 
steady wind model \citep{chandra+12b,chandra+15,dwarkadas+12}. Therefore, we generalize our model and keep $s$ as a free parameter.

The  radio synchrotron  emission undergoes suppression,   due to free-free absorption (FFA) by the ionised CSM \citep{chevalier82a}
and/or due to   synchrotron self-absorption  (SSA) by the same electron population
responsible for the radio  emission \citep{chevalier98}.
In a model where  FFA is the dominant absorption mechanism, the radio flux density, $F(\nu,t)$, can be expressed  as  \citep{chevalier82a,weiler+02}
\begin{equation}
F(\nu,t)=K_{1} \left( \frac{\nu}{5\hspace{0.1 cm} \rm GHz}\right)^{-\alpha} \left( \frac{t}{1000\hspace{0.1 cm} \rm day}\right)^{-\beta}e^{-\tau_{\rm ffa} (\nu, t)},
\label{eq:ffa1}
\end{equation}
where $K_{1}$ is a normalization factor whose value is equal to the radio flux density at 5\,GHz measured at day 1000 after the explosion. The indices $\alpha$ and $\beta$ denote the spectral and temporal evolution in the optically thin phase, respectively. The radio spectral index $\alpha$ is related to the electron energy index $p$ (in $N(E)dE\propto E^{-p}dE$) as $\alpha=(p-1)/2$. 
The assumptions that the energy density in the particles and the  magnetic field are proportional to the postshock energy density lead to the parameter $\beta=(3+\alpha)(ms+2-2m)/2-3m$.
Here $\tau_{\rm ffa}$ is the  FFA optical depth   due to  ionized CSM external to the emitting material, and can be written as 
 
\begin{equation}
\tau_{\rm ffa}(\nu,t)=K_{2}\left(\frac{\nu}{5\hspace{0.1 cm}\rm GHz}\right)^{-2.1}\left(\frac{t}{1000\hspace{0.1 cm} \rm  day}\right)^{-\delta},
\label{eq:ffa2}
\end{equation}
where $K_{2}$ denotes the free-free optical depth at 5 GHz measured at day 1000 after the explosion. As the shock wave expands, the optical depth decreases with time as $t^{-\delta}$, where $\delta$ is related to shock deceleration parameter $m$ as  $\delta=m(2s-1)$.

 The high densities in SNe IIn indicate that FFA is likely to be the dominant absorption mechanism \citep{chevalier98}.
 However, since \snem\ initially exploded as a Type Ib/c supernova, SSA could be important and we account for this possibility.
The radio flux density for the SSA dominated synchrotron emission can be written as \citep{chevalier98}:
\begin{equation}
F(\nu,t)=K_{1} \left( \frac{\nu}{5\hspace{0.1 cm} \rm GHz}\right)^{2.5} \left( \frac{t}{1000\hspace{0.1 cm} \rm day}\right)^{\beta'} (1- e^{-\tau_{\rm ssa}(\nu, t)}),
\label{eq:ssa1}
\end{equation}
where the optical depth is characterised by SSA due to the relativistic electrons at the forward shock. The SSA optical depth $\tau_{ssa}$ is given by
\begin{equation}
\tau_{\rm ssa}(\nu, t)=K'\left(\frac{\nu}{\rm 5\hspace{0.1 cm} GHz}\right)^{-(\alpha+2.5)}\left(\frac{t}{1000\hspace{0.1 cm}  \rm day}\right)^{-(\beta'+\beta)}.
\label{eq:ssa2}
\end{equation}
$K_{1}$ and $K'$ are the flux density and optical depth normalisation factors similar to the case of FFA. The flux density in the optically thick and
thin phases behave as $\nu^{2.5}$ and $\nu^{-\alpha}$, respectively. 
Here $t^{-\beta}$ is the time evolution of the  flux density in the optically  thin phase, and
$t^{\beta'}$   for the optically thick  phase.
The exact form 
of $\beta$ and  $\beta'$  depend upon the 
scalings of magnetic field, $B$,  and electron energy density \citep{chevalier96}. If we assume that the magnetic energy density and 
relativistic electron energy density scale with post shock energy density \citep[model 1 of ][]{chevalier96}, then  the
optically thick light curve,
$F_\nu(t) \propto R^2B^{-1/2}$,  leads to  
$\beta'=2m+0.5$ and  $\beta=(p+5-6m)/2$ for $s=2$. 

In the high density environment of  SNe IIn, the reverse shock becomes radiative and a cool dense shell  is likely to form between the forward and reverse shocks because of the higher density of the CSM. In this case  a fraction of the cool gas mixed into the synchrotron emitting region can
 cause  internal FFA. \citet{weiler+90} developed a formulation for  internal FFA to explain the radio data of SN 1986J.
 The early radio emission is described by the escape probability from a clumpy region.
\citet{chandra+12b} showed that a modest amount of cool gas mixed into the synchrotron emitting gas was sufficient to explain the absorption in
SN 2006jd radio emission.
 The internal FFA model is described as \citep{weiler+90}:
\begin{eqnarray}
F(\nu,t)=K_1 
\left(\frac{\nu}{5 \,{\rm GHz}}\right)^{-\alpha}
\left(\frac{t}{1000\, {\rm day}}
\right)^{-\beta} 
\left(  \frac{1-\exp(-\tau_{\rm intFFA}(\nu,t))}{\tau_{\rm intFFA}(\nu,t)}\right), \nonumber \\ 
\tau_{\rm intFFA}(\nu,t)=K_3 \left(\frac{\nu}{5 \,{\rm GHz}}\right)^{-2.1}
\left(\frac{t}{1000\, {\rm day}} \right)^{-\delta^{\prime}}, 
\label{eq:int}
\end{eqnarray}
where  we allow $s\ne 2$; $\alpha$ and  $\beta$ are
expressed as above for the FFA model.
\citet{weiler+90} have assumed  that the internal absorbing gas is homologously expanding
with density $n\propto R^{-3}\propto t^{-3m}$, which will lead to  $\tau_{\rm intFFA}\propto n^2R\propto t^{-5m}$, or $\delta^{\prime}=-5m$.
However, this may not be the case and the prevailing conditions   are unknown. We account for 
 other possibilities for  $\delta^{\prime}$ and allow it  to be a free parameter.

 To apply the CSM interaction model to  \snem\  radio  data, we consider the same sequence of events as  given by  \citet{cc06}.
\snem\ lost all of the H-envelope  before the  explosion and exploded as a stripped envelope supernova.  However, 
 the H-rich material did not have enough time to be completely dispersed into the interstellar medium and was located close enough so that  the  ejecta were able to catch up with this H-shell, and the subsequent interaction produced radio and X-ray emission and  broad H$\alpha$ emission. The H-envelope may have been
emitted during a red supergiant phase with a wind speed of $v_{\rm w}\sim 10$\,\kms, and got accelerated by the fast W-R winds to $30-50$\,\kms.

\begin{figure*}
\centering
\includegraphics*[angle=0, width=0.98\textwidth]{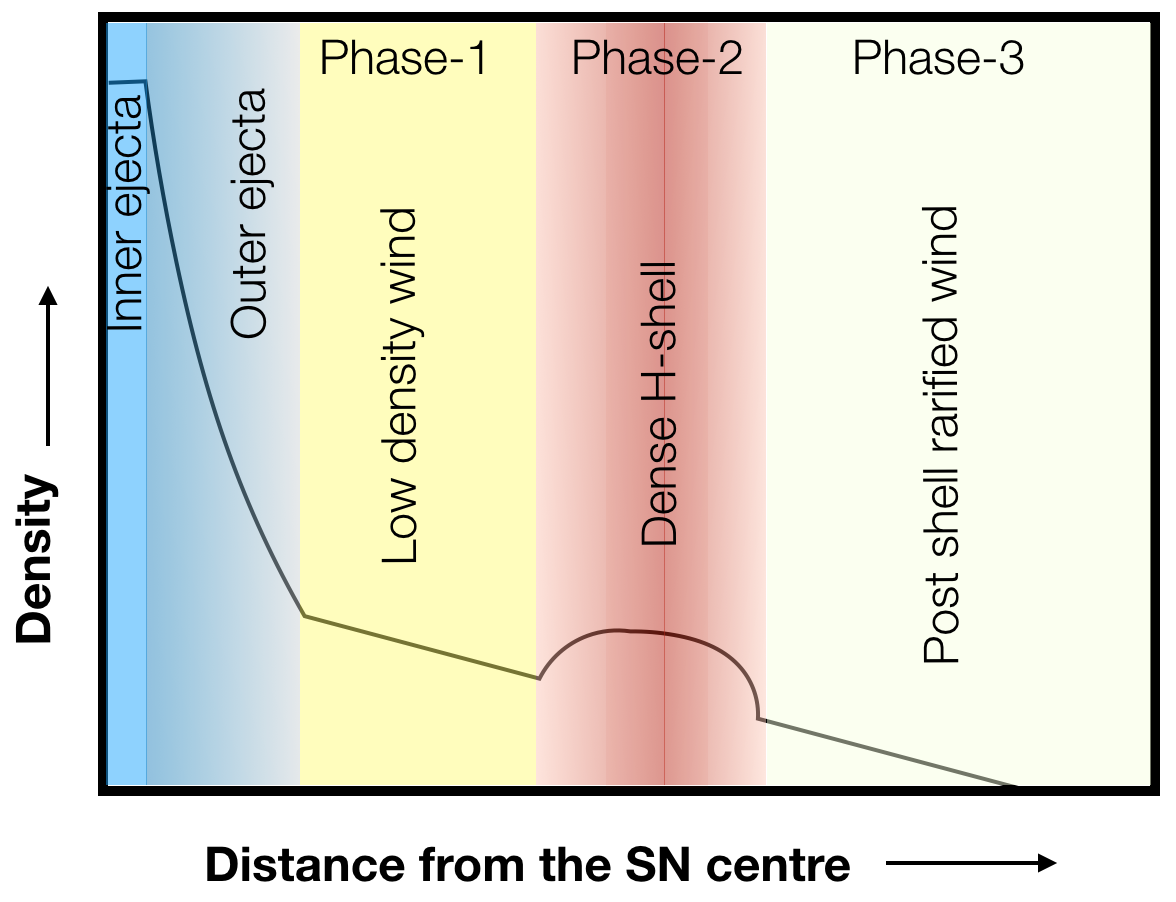}
\caption{Cartoon diagram of various ejecta and wind density profiles. \label{phase}}
\end{figure*}

Several phases of evolution can be identified, depending on the circumstellar density profile, which is illustrated in Fig. \ref{phase}.
Initially (Phase - 1), the outer supernova ejecta interact with the fast wind of the progenitor star, followed by interaction with shocked wind.
This phase of low density interaction was missed in radio and X-ray observations of SN 2001em.
Next is a transition between Phase - 1 and Phase  -2 during which the increasing pressure of the shocked ejecta drives a shock front into
the dense shell     \citep{cl89}.  The increasing pressure is accompanied by increasing X-ray and radio emission.  In Phase - 2, the emission is
mainly from the shell interaction.  This emission drops when the shock front comes out of the dense region and eventually the interaction
with the outer circumstellar medium dominates the emission (Phase - 3).  The transition may occur over several dynamical timescales during which the
shock velocity is roughly constant \citep{vanmarle+10}
 or might accelerate \citep{harris+16}.

\subsection{Model fits to radio data}

We now carry out detailed fits to the radio data. We use   Markov chain Monte Carlo (MCMC) fitting  using the Python package emcee \citep{fm+13}. 
We choose 200 walkers, 1000 steps, and flat priors on all of the parameters.
We obtain best-fit values
(at 68\% confidence interval, i.e., $1\sigma$).

\snem\ lacks early radio and X-ray data and its Phase--1 was completely missed. We need to fit for Phase--2 and Phase--3. Since the data are sparse at later epochs, it is not clear when   the  Phase--2 ended, although towards the end of this section we try to answer this question.    
We start with global  fits using  the FFA and  the SSA models to all the data. We list the best fit parameters of FFA and SSA models in columns 1 and 2 of Table \ref{tab:para-gmrt}, respectively. We note that none of the models are a good fit to the data with  large  values of reduced Chi-square ($\chi_\nu^2$). In addition,  both models give best-fit $\alpha < 0.5$, suggesting a hard electron energy index $p < 2$, which is observationally not the case (Fig. \ref{fig:radiodata}).
While the FFA fit is slightly better, the derived best values of $\alpha$, $\beta$ and
$\delta$ result in  
$m\approx 1.46$, which is not consistent with the hydrodynamic model.

One can independently estimate the significance of SSA model based on the velocity of the ejecta at the SSA peak \citep[][ and references therein]{cf17}.
The 8.5 GHz peak comes around day 1000 with a peak flux density of $\sim 1.7$ mJy.  Using Eq. 59 of \citet{cf17}, we  obtain the size deduced from the  SSA model as $\sim 1\times10^{16}$ cm, suggesting  the velocity $\sim1000$\,\kms\, at 1000 days. This is  quite small for a SN shock and also smaller than the
ejecta velocity of  $>1800$\,\kms\, around day 970 obtained from the H$\alpha$ FWHM   \citep{soderberg+04}, suggesting SSA is not the dominant radio absorption.

\begin{deluxetable}{|l|l|l|l|l|}
\tablecaption{Best fit   parameters  for the  FFA \& SSA models to the radio data \label{tab:para-gmrt} }
\tablehead{
\colhead{FFA} & \colhead{SSA} & \colhead{Int. FFA} &  \colhead{Ext.$+$Int. FFA} &  \colhead{Ext.$+$Int. FFA ($s=2$)} 
}
 \colnumbers
\startdata
$K_{1}=2.43^{+0.11}_{-0.10}$     &  $K_1=3.11^{+0.20}_{-0.19}$ & $K_{1}=2.79^{+0.16}_{-0.14}$ &          $K_1=3.07^{+0.16}_{-0.15}$  &           $K_1=3.00^{+0.17}_{-0.16}$  \\
$K_{2}=   0.41^{+0.04}_{-0.03} $ &  $K'=0.75^{+0.08}_{-0.07}$ &    $K_{3}=1.37^{+0.16}_{-0.14}$ &    $K_2=0.13^{+0.02}_{-0.02}$ &  $K_2=0.15^{+0.02}_{-0.02}$   \\
$\alpha=0.41^{+0.03}_{-0.03}$   &  $\alpha=0.40^{+0.03}_{-0.03}$ & $\alpha=0.48^{+0.04}_{-0.04}$ &     $K_3=0.68^{+0.13}_{-0.11}$  &    $K_3=0.61^{+0.15}_{-0.13}$  \\
$\beta=1.49^{+0.07}_{-0.07}$     & $\beta=1.45^{+0.07}_{-0.07}$ &   $\beta= 1.68^{+0.09}_{-0.09}$ &   $\alpha=0.69^{+0.05}_{-0.05}$  &    $\alpha=0.65^{+0.05}_{-0.05}$  \\
$\delta=7.28^{+0.30}_{-0.29}$.   &  $\beta^{\prime}=8.39^{+0.31}_{-0.31}$ &  $\delta^{\prime}=7.04^{+0.33}_{-0.33}$ &    $\beta=1.63^{+0.08}_{-0.08}$ &  $\delta=2.02^{+0.07}_{-0.07}$   \\
& & &$\delta=1.60^{+0.16}_{-0.08}$    & $\delta^{\prime}=10.48^{+0.89}_{-0.81}$   \\
& &  &  $\delta'=10.02^{+0.58}_{-0.64}$    & \\
\hline
$\chi_{\nu}^{2}=3.27$  & $\chi_{\nu}^{2}=3.28$ &   $\chi_{\nu}^{2}=2.89$ &          $\chi_{\nu}^{2}=2.19$  & $\chi_{\nu}^{2}=2.21$   \\
{\it d.o.f.}=56  & {\it d.o.f.}=60   &     {\it d.o.f.}=56        & {\it d.o.f.}=54  &  {\it d.o.f.}=55  \\
\enddata
\end{deluxetable}

\begin{figure*}
\centering
\includegraphics*[angle=0, width=0.98\textwidth]{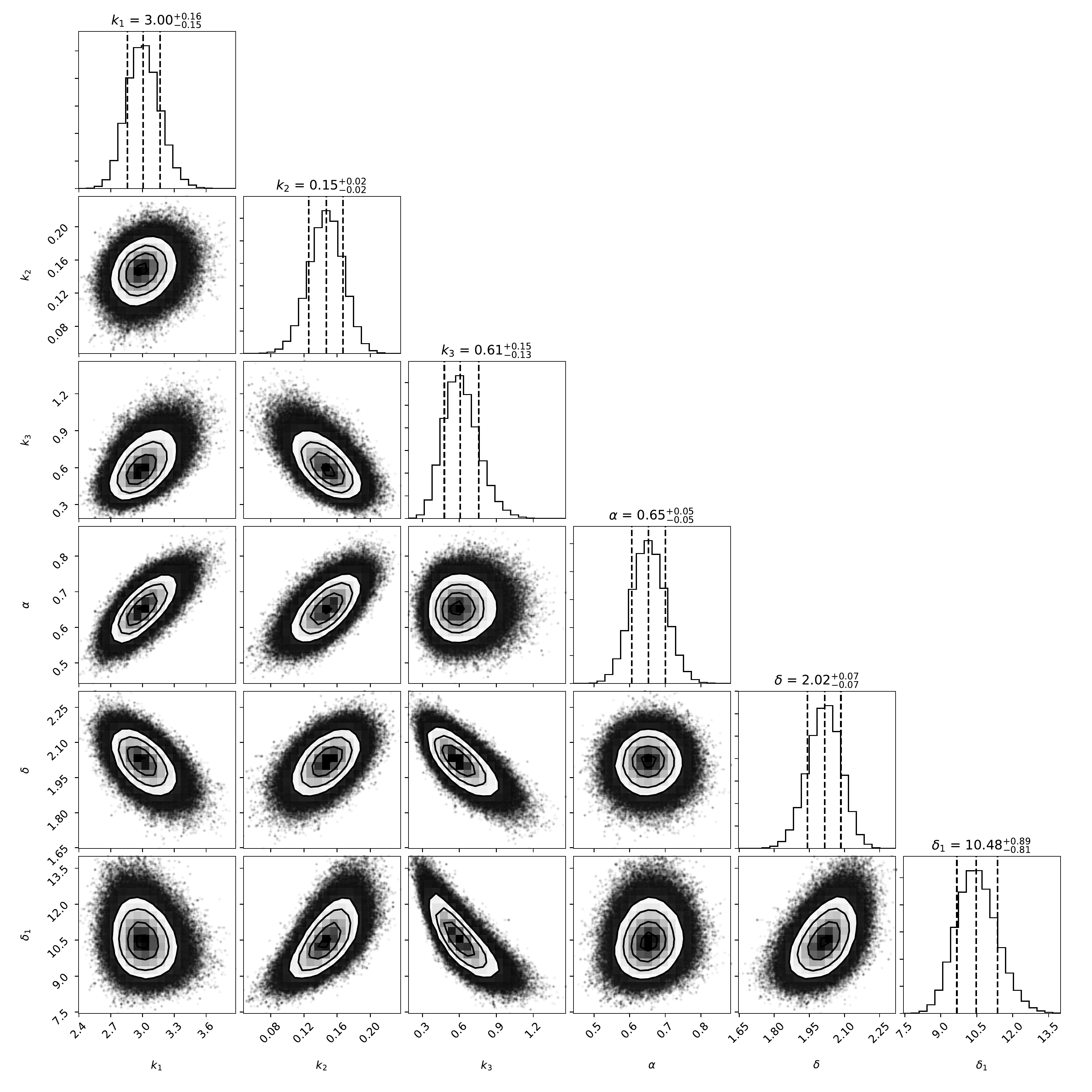}
\caption{Corner plot with the MCMC fit to the internal$+$external model of the radio emission with $s=2$  for SN 2001em \label{fig:corner}.}
\end{figure*}

\begin{figure*}
\centering
\includegraphics*[angle=0, width=0.9\textwidth]{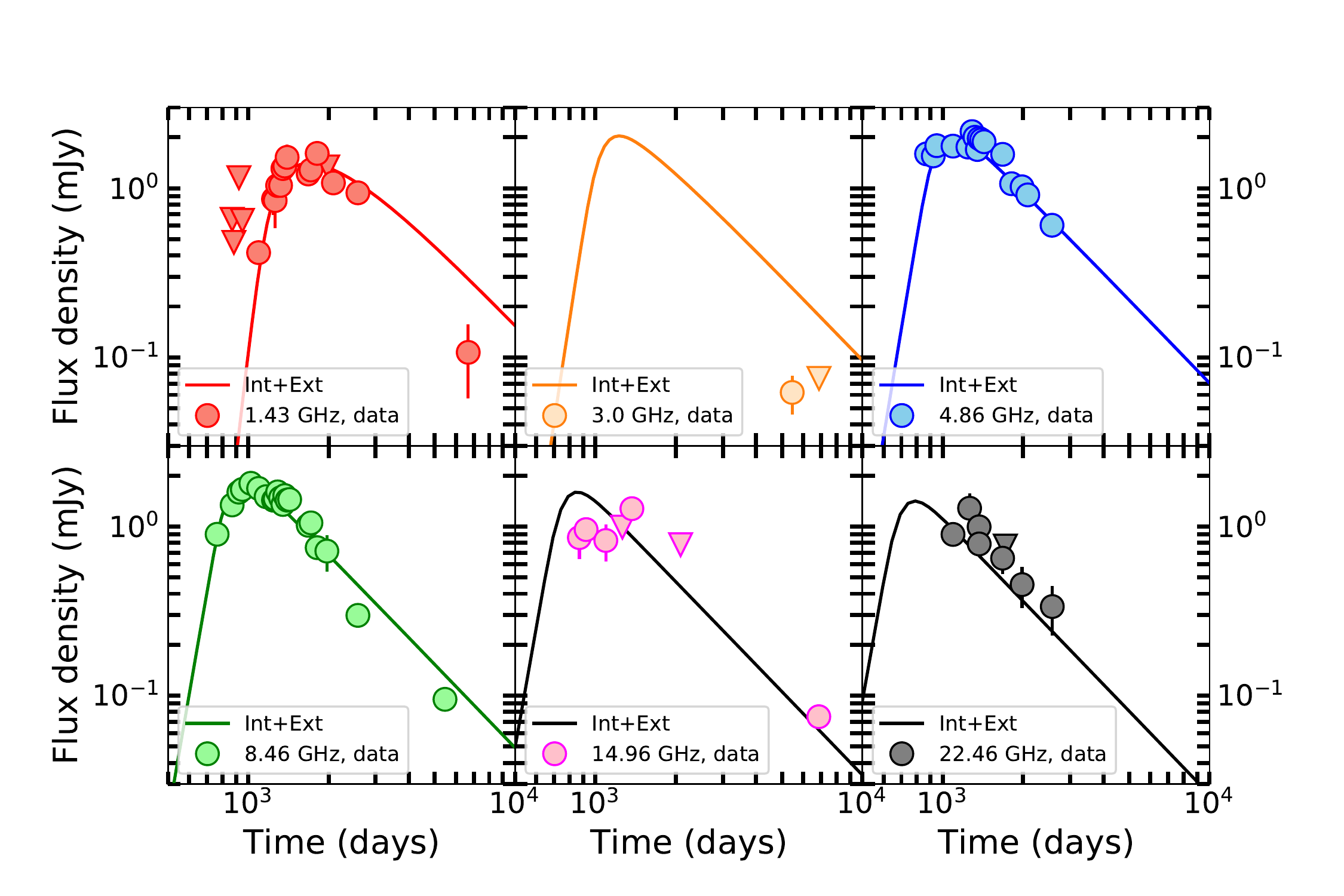}
\vspace*{-1cm}
\includegraphics*[angle=0, width=0.9\textwidth]{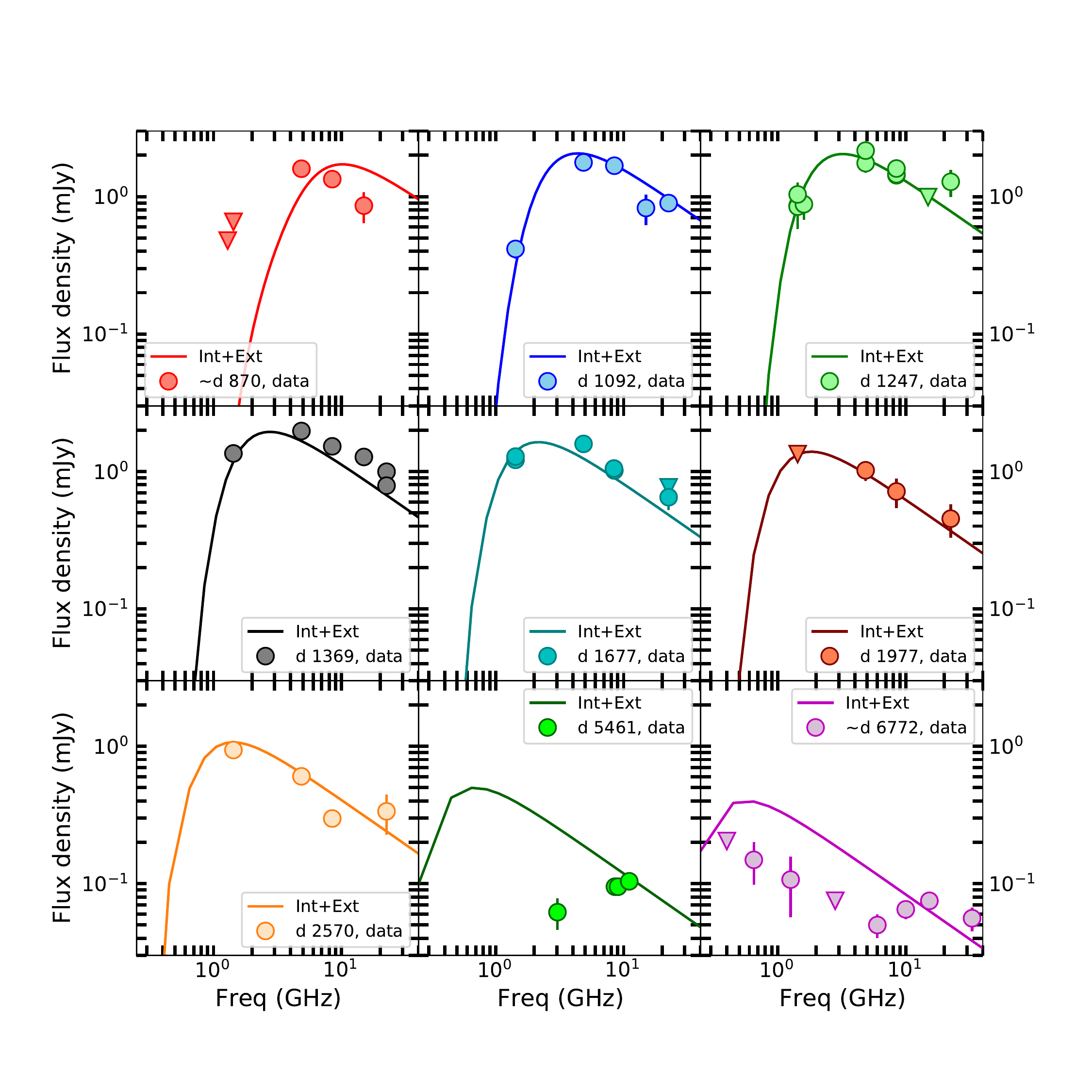}
\caption{Radio light curves (top panel) and spectra of SN 2001em fit with the internal$+$external FFA model.   The model fits the data well  between 1000 -- 2500 days. 
\label{fig:sn2001em-lfit}}
\end{figure*}

The internal FFA model of \citet{weiler+90} is  a better representation of the data with improved fit statistics (column 3 of Table \ref{tab:para-gmrt}).  
As in the case of SN 1986J, we now account for the possibility that the radio emission is absorbed by both external as well as internal
thermal absorbers. 
 We note that internal$+$external FFA gives better fits (column 4 in Table \ref{tab:para-gmrt}). The derived value of 
$s$ for this model is $s=1.80\pm0.26$. This value is very close to $s=2$, so we again fit the internal$+$external FFA  model, but
 fixing $s=2$,  so that $\beta=3+\alpha-\delta$. We also exclude data points from day 5460 onwards as they appear  to arise from a different
component (\S \ref{sec:visual}). This model 
(column 5 in Table \ref{tab:para-gmrt}) provides a good fit to the data amongst all the models
considered. 
This model  results in  $m=0.67\pm0.08$, probably indicating  deceleration after the shock entered the H-shell. Figure \ref{fig:corner} shows the corner plot with the results of the MCMC fit for this model. The  parameters are well constrained with internal absorption being the dominant absorption mechanism. 
  In Fig. \ref{fig:sn2001em-lfit}, we plot the best fit light curves and spectra. 
  
We note that the reduced  $\chi_\nu^2$ is still  large. This is probably because the evolution is  not smooth and there is  detailed structure in the radio light curves. The contribution to the $\chi^2$ is more from the radio fluctuations  rather than the model fits.  The contribution of internal absorption
is 4 times higher than the FFA. This means that the radio emission is arising from a shocked region containing clumpy gas.
The clumpiness could come from hydrodynamic instabilities at the contact discontinuity or from clumpiness in the ejecta or the surrounding region.

 In Fig. \ref{fig:sn2001em-lfit}, we plot light curves and spectra along with the best fit model. 
 For the late time spectrum, we combine the high frequency VLA data around day 6870 with the  low frequency
upgraded GMRT data on day 6675  and plot them at an average epoch of $\sim 6772$ day.
 The model fits adequately represent the  light curves   at early times and capture the overall evolution rather well.
 However, the fits to  the spectra are not as good as the light curves. 
 There are two points to note: 
on day 870, the spectrum seems to suggest a peak at frequencies lower than predicted by the model.
This could mean the actual density is lower than that predicted by the  model.
One explanation is  if the shock entered not so long ago in the dense shell, and the shock is 
 still catching up to the high density region within the shell, which probably is somewhere at the center part (Fig. \ref{phase}). However, the spectrum consists of only three measurements and  fluctuations might simply be caused by the
 clumpy density profile. 
 There is one early data point on day 764 at 8.46 GHz which is consistent with the overall light curve model (Fig. \ref{fig:sn2001em-lfit}). 
  
   The most  interesting trend is seen at the latest near-simultaneous spectra on day 5460 and 6772 (Fig. \ref{fig:sn2001em-lfit}). While the model predicts an optically thin spectrum at all frequencies, the observed spectrum turns over at around $\ge 3$ GHz
and becomes optically thick. 
 The X-ray observations at this epoch result in an upper limit, which is not constraining.
The behavior can be explained if there is a central absorbed component, which is unveiled as the absorption of the inner source  and the flux of the outer source decline.
The lower than expected optically thin spectrum at lower frequencies could be an indication of the forward
 shock moving in a  low density medium.
 This can also explain the spectrum on day 2500. The 8.5 GHz flux is  smaller than expected from the best fit model, which could be an indication that the shock is leaving the dense shell and moving into the
 steeply decreasing density, consistent with the X-ray data.  The higher flux at 22.5 GHz component could then be coming from the  central absorbed component, seen  clearly from
 day 5461 onwards.
 
\begin{figure*}
\centering
\includegraphics*[angle=0, width=0.98\textwidth]{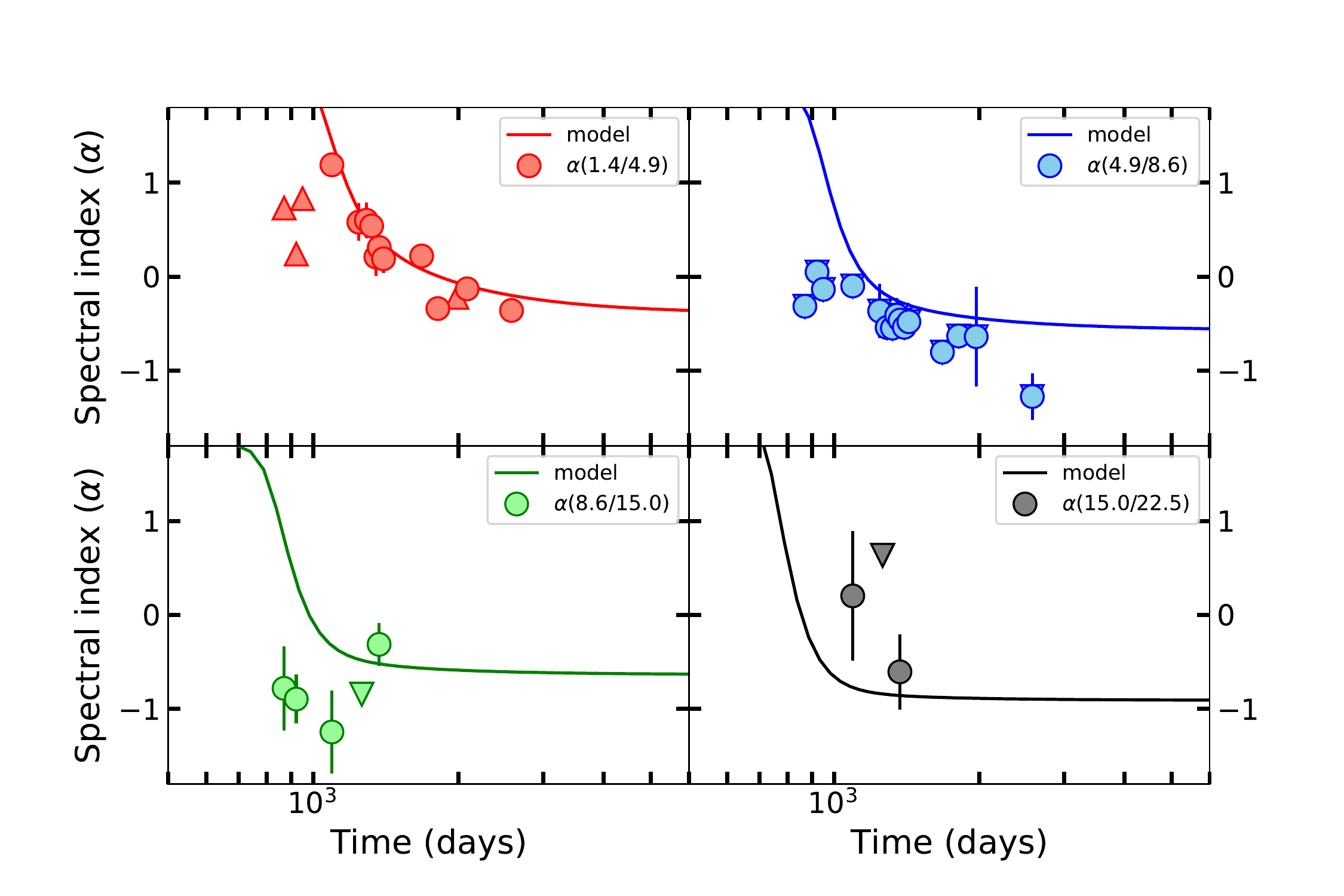}
\caption{Spectral index at different adjacent frequencies. The spectral index  moves from positive to negative values
indicating a transition from optically thick phase to optically thin phase. The thick solid lines are the best fit internal$+$external absorption models. The triangles are   
limits, in the direction that the triangle is pointing. \label{fig:sn2001em-alpha}}
\end{figure*}

\begin{figure}
  \centering
\includegraphics*[angle=0, width=0.8\textwidth]{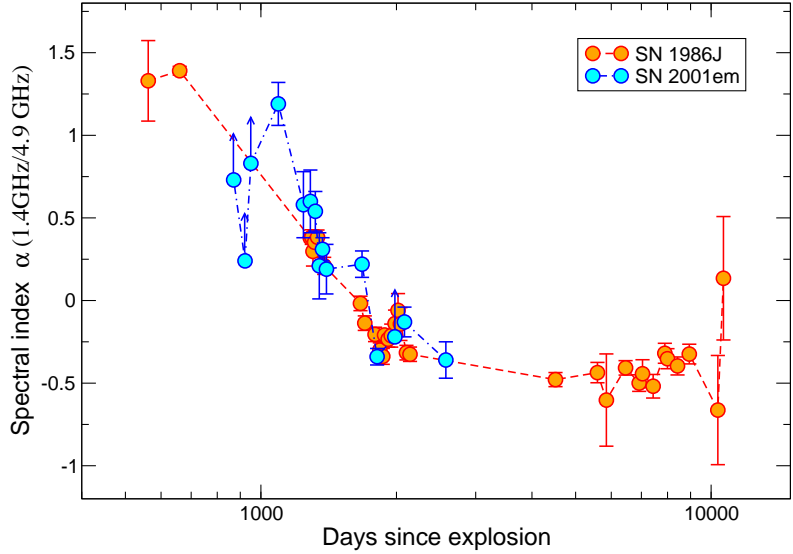}
\caption{The spectral index evolution for SN 2001em and SN 1986J between 1.4 and 4.9 GHz frequencies. The SN 2001em spectral evolution approximately follows that of SN 1986J. The arrows represent the  limits in the direction they are pointing. \label{fig:alpha86j}}
\end{figure}

 In Fig. \ref{fig:sn2001em-alpha}, we plot the spectral indices ($\alpha$;   in $F(\nu) \propto \nu^\alpha$) between the observed adjacent frequencies.  The values of $\alpha$ are mostly negative
after around 1000 days indicating an optically thin medium, except between 1.4 and 4.9\,GHz.
The values of $\alpha$ between these two frequencies are positive up to the first 3.5 years before making a transition to the optically thin phase. 
We derive the model spectral indices from the best fit parameters and overplot them in Fig. \ref{fig:sn2001em-alpha}. They  seem to represent the data reasonably well,  giving support to the  internal$+$external model.  
We compare the spectral evolution for SN 2001em   to that of SN 1986J, which was also 
explained by the  internal$+$external absorption model. The resemblance is striking (Fig. \ref{fig:alpha86j}), suggesting that
the presence of a cool-dense shell or clumps in the case of dense interaction is  common and contributes towards extra absorption of radio emission as internal FFA.

\subsection{Mass-loss rate}
\label{sec:massloss}

We can now attempt to constrain the mass-loss rate from the radio data.
\citet{weiler+02} and \citet{cf17} derived the  mass-loss rate  formula considering multiple absorption components:
\begin{equation}
\left(\frac{\dot M_{-3}}{v_{w1}}\right)=7.5\times10^{-2}  \left( \frac{\left< \tau_{\rm eff}\right>}{1+[n({\rm He})/n({\rm H})]}\right) ^{0.5}
 \left(\frac{\nu}{\rm GHz}\right)^{1.06}  \left(\frac{T}{10^4\,\rm K}\right)^{0.67}
  \left(\frac{v_s}{10^4\,{\rm km\,s^{-1}}}\right)^{1.5} \left(\frac{t}{1000\,\rm d}\right)^{1.5}.
\end{equation}
Here, $\left<\tau_{\rm eff}\right>$ is the effective optical depth considering all absorption effects. 
We need the velocity of the radio emitting shock  $v_s$ and $T$,  the CSM temperature.
We assume $T\sim10^4$\,K.
 To estimate the velocity, we note that the H$\alpha$ emission had a  FWHM  of $\sim 
 $1800\,\kms. 
 As the supernova at the time of explosion was a stripped envelope supernova, 
one expects  H$\alpha$ to arise from the shocked CSM. In such a case the H$\alpha$ width reflects the lower limit to the
forward shock velocity, the actual velocity will be closer to the base of the H$\alpha$  profile. However, it gives 
the order of the velocity.
From X-ray observations, the temperature is 
$\ge 15 $\,keV, corresponding to a velocity  $>3500$\,\kms\ velocity. 
From VLBI observations, \citet{schinzel+08} put an upper limit to the expansion velocity of 6000\,\kms. 
\citet{bietenholz+05} obtained an expansion
velocity of $5800\pm10000$\,\kms\ in their measurements on day 1770,  which is consistent with an expansion velocity from zero to $10^4$\,\kms.
In the case of SN 2014C, \nustar\ data were available and the temperature was found to be at $\sim 20$ keV \citep{margutti+17}. 
Similarly for SN 2010jl, the \nustar\ measurements revealed the forward shock temperature to be $\sim 18$ keV \citep{chandra+15},
corresponding to a velocity of $\sim 3700$ \kms.
In case of SN 1996cr, which is also interacting with a dense shell, \citet{dwarkadas+10} found the X-ray emitting shock to be moving with 
$4740$\,\kms.
 VLBI observations of SN 1986J
showed an expansion speed around 1000 days to be $5700$\,\kms \citep{bietenholz+04}.  
 We thus adopt 4000\,\kms\ to be an expansion speed for SN 2001em. 
 
Our best fit model predicts mass-loss rates of $(5.2-0.9)\times10^{-3}$\,\mdot\ during epochs 170 to 370 years before explosion, for a wind speed of 
50\,\kms.

\subsection{X-ray analysis}
\label{sec:xray}

\begin{figure*}
\centering
\includegraphics*[angle=0, width=0.65\textwidth]{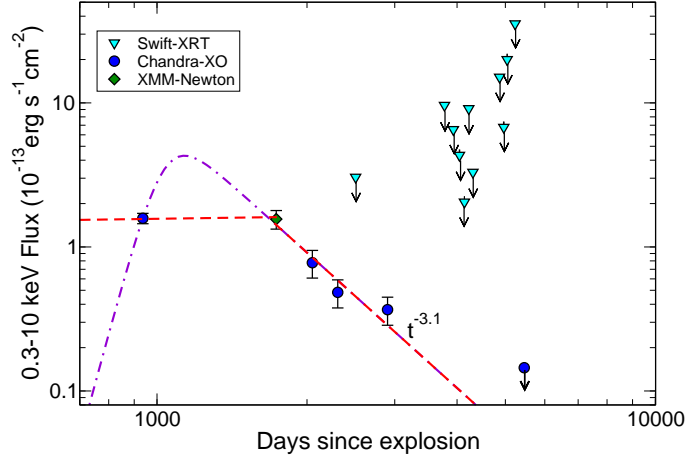}
\caption{ The 0.3--10.0\,keV X-ray light curve of SN 2001em. 
Due to only two data points between 934 and 1735 days,  the X-ray evolution is poorly constrained before day 1735, except that the 
X-ray peak lies somewhere between 934 and 1735 days. One possibility 
is that the X-ray flux remained constant between these two epochs  (red dashed line) and 
  then started to fall rapidly.  However, various evolutions  are possible with a peak between these two epochs. 
  The purple dot-dashed line shows one such possibility for illustrative purposes. The evolution after 1735 days is well constrained,
  with the flux declining as $F_{\rm Xray}(t) \propto t^{-3.1}$.  .
\label{fig:sn2001em-xray}}
\end{figure*}

The X-ray data cover  epochs ranging from day 934 to day 5454.  While the  radio data are sampled  more densely between day 764 and day 2575,  the X-ray data have sparser coverage up to day 1735 and have better  coverage afterwards  (Fig. \ref{fig:sn2001em-xray}).
However, none of the \swift\ observations (which were taken at  later epochs) resulted in  a detection and the upper limits are not constraining. 
  The last epoch \chandra\  observation on day 5464 also resulted in an upper limit.
   The  highest 
  X-ray flux corresponds to  an unabsorbed $0.3-10$\,keV X-ray luminosity  $L_{\rm Xray}= 1.25\times10^{41}$\,\ergs, placing it among the brightest X-ray supernovae  \citep{chandra18}.

 The \xmm\, data at day $\sim 1735$  are the best dataset  to constrain the temperature. 
We fit the data with an absorbed thermal bremsstrahlung model. We find that the temperature in our model is not well constrained. Within the sensitivity of the \xmm\, energy range,
it is not bounded by an upper limit. The   lower bound is $13$\,keV.  Thus we fix the temperature to 20 keV to carry out the X-ray analysis.  \citet{pooley+04} found a temperature of $80$ keV,  but 
due to the limited \chandra\, band it is not possible to constrain the temperature to these high values. Our value is assumed to be like that found in SN 2014C  by \citet{margutti+17}, who
found the
plasma temperature producing X-rays to be $\sim 20$\,keV  using {\it NuSTAR} observations.

 The temporal evolution between days 934 and 1735 is defined by 2 data points, giving  $F_{\rm Xray}(t)  \propto t^{0.05}$ up to around the \xmm\, epoch, i.e. day 1735.  One possibility 
is that the X-ray flux remained constant between these two epochs and 
  then started to fall rapidly afterwards.  However,   a peak between these two epochs is plausible. We plot one such
  possibility in Fig. \ref{fig:sn2001em-xray}.  Due to the lack of early data points, it is not possible to further constrain the evolution.

The light curve falls steeply after day 1735,    steepening to   $F_{\rm Xray}(t) \propto t^{-3.12\pm0.66}$.
Such an evolution at late time is possible if the forward shock wave  moved out of the shell and into lower density gas.
In the  radio spectrum on   $\sim 2500$ d, the 8.5 GHz data point is much below the standard model with $s=2$,  consistent with the
X-ray prediction.  
There is an increase in the 22 GHz flux density, which can be explained if this component is coming from a central absorbed component, causing the flux density to rise.

Since the temperature of the emitting region is higher than can be measured in the 0.3--10 keV bandpass of the X-ray detectors used, our values represent the spectral flux
rather than the total X-ray flux.  
For bremsstrahlung emission and slow cooling, the X-ray spectral luminosity  can be written as
 $L_{\rm Xray} \propto n^2 V/T^{1/2}$, where $V$ is the emitting volume \citep{chandra+12b}. As $R \propto t^{-m}$
 and $n \propto R^{-s}$, this implies $L_{\rm Xray} \propto t^{2m(1-s)+1}$.  A value of $s\sim 3$ can reproduce the steeper decline.

 Between day 934 and day 2900, the column density evolves from $\sim (2.7\pm0.6) \times 10^{21}$\,\nh\, to   $(\sim 1.3\pm0.7) \times 10^{21}$\,\nh\, 
 (Fig. \ref{fig:xmod}D).  Afterwards we do not have sensitive measurements to constrain the column density. While the error-bars are large, we attempt to fit a power law to the evolution of column-density and it fits best with an index of $-0.71\pm0.23$ (Fig. \ref{fig:xmod}D).
 Thus, there is  evidence for evolution of the column density to a value  higher than the Galactic column density
$N_H=6\times10^{20}$\nh \citep{spitzer78}, corresponding to the reddening toward SN 2001em,  $E(B-V) = 0.1$ \citep{schlegel+98}.

\subsection{Origin of X-ray emission}

A critical question is to find the origin of X-ray emission.  \citet{cc06} modelled the origin of the radio and X-ray emission of \snem\  to be an 80\,keV region produced by a
 5500 \kms\ reverse shock. However, their  modelling had two issues:  1) the velocity of the forward shock was assumed to be
1800 \kms\ based on H$\alpha$ measurements \citep{soderberg+04} and, 2) X-ray emitting plasma was considered to have a 
  temperature of 80\,keV based on  the results of   \citet{pooley+04}. 
We note that the speed of 1800\,\kms\ represents the FWHM of the
  H$\alpha$ profile which is likely to be a lower limit on the forward shock velocity and, as we argue above (\S \ref{sec:massloss}), the forward
  shock velocity is $\sim 4000$\,\kms. The shock temperature of 80\,keV was measured from \chandra\ observations with a limited 0.3--10\,keV band, which is not sensitive to such 
  high temperatures. As discussed above, more robust temperature measurements have been done for a couple of supernovae in a combined \xmm\ and \nustar\
  spectral analysis covering an energy range 0.3--80\,keV.

The forward shock origin of X-ray emission  is further supported by two arguments. In supernovae interacting with dense medium, it is  likely that the reverse shock  becomes radiative  and
a cool dense shell forms. Our radio modeling gives  evidence of the formation of a cool dense shell, hence that the reverse shock is
radiative. We estimate the cooling time for reverse as well as forward shocks for $n=9$. 
For the mass-loss rates estimates at ages 773 to 1670 days (\S \ref{sec:massloss}), the ratio of cooling  time
to expansion time is $\le 1$ for most of the period for the reverse shock, whereas, for forward shock it is around 20--200. This further confirms that, while the reverse shock is
radiative, the forward shock is adiabatic. 
Radiative cooling is more important at the reverse shock and
the cool dense shell is likely to absorb much of the X-rays emanating from the reverse shock.

We find that the column density is marginally evolving and there is an excess of column density of  $\sim (2.1-0.7) \times 10^{21}$\,\nh. 
Since the X-ray emission is coming from the forward shock, this column density represents the column density of the CSM.
The evolution of CSM column density 
 means that CSM was  not completely ionized between the first two X-ray epochs and photoabsorption is important. 
\citet{ci12}
have estimated that if a supernova is in the cooling regime, a 10,000 \kms\ shock wave is capable of completely ionizing the surrounding medium, but a 5000 \kms\  shock wave is not.
The state of ionization in the circumstellar gas is characterized by the value of the ionization parameter $\zeta$.  For a hydrogen helium plasma: 
\begin{equation}
\zeta \equiv \frac{L_{CS}}{r^2n}=42  \left(\frac{L_{\rm CS}}{10^{41}\,\rm erg\,\rm s^{-1}}\right)
\left(\frac{\dot M_{-3}}{v_{w1}}\right)^{-1}.
\end{equation}
We estimate this value for the epochs at which we have estimated the mass-loss rate from radio data. During these epochs, the X-ray luminosity is nearly constant and the parameter $\zeta$ varies between $\sim 50-200$.
It is  a regime where the CNO elements may be completely ionized, but Fe is not \citep{hatchett+76}.  Hence the evolving column density can be
explained by not fully-ionized CSM plasma.

Our radio measurements imply a CSM column density of $( \rm a\, few) \times10^{22}$\,cm$^{-2}$, which is nearly an order of magnitude larger than
our best fit X-ray measurements. This trend has been seen in other supernovae as well, and has been attributed to the asymmetric nature of the CSM
and/or ejecta \citep{chandra+12b}. 
An asymmetrical density distribution  is a natural outcome of the common envelope scenario.
 \citet{sophie+20}  have carried out hydrodynamic modeling of binary coalescence and found a toroidal distribution of
 the extended CSM in the binary's equatorial plane.

Due to the low absorption in X-ray and
 radio emission, \citet{cc06}  invoked clumpiness in the absorption, but  emission  by a smooth CS shell. 
 As ejecta enter into the shell, clumps get fragmented and
 mixed in the forward shock depositing their full energy to the shock. 
 In this model, absorption is  dominated by the clumpy medium but the emission  is dominated by the smooth shocked medium.
This model predicts a  soft component absorbed and reprocessed into optical/UV, in addition to a hard component from the hot forward shock.
 The lack of data in these bands does not allow us to test this possibility.

 \begin{figure}
 \centering
\includegraphics*[width=0.49\textwidth]{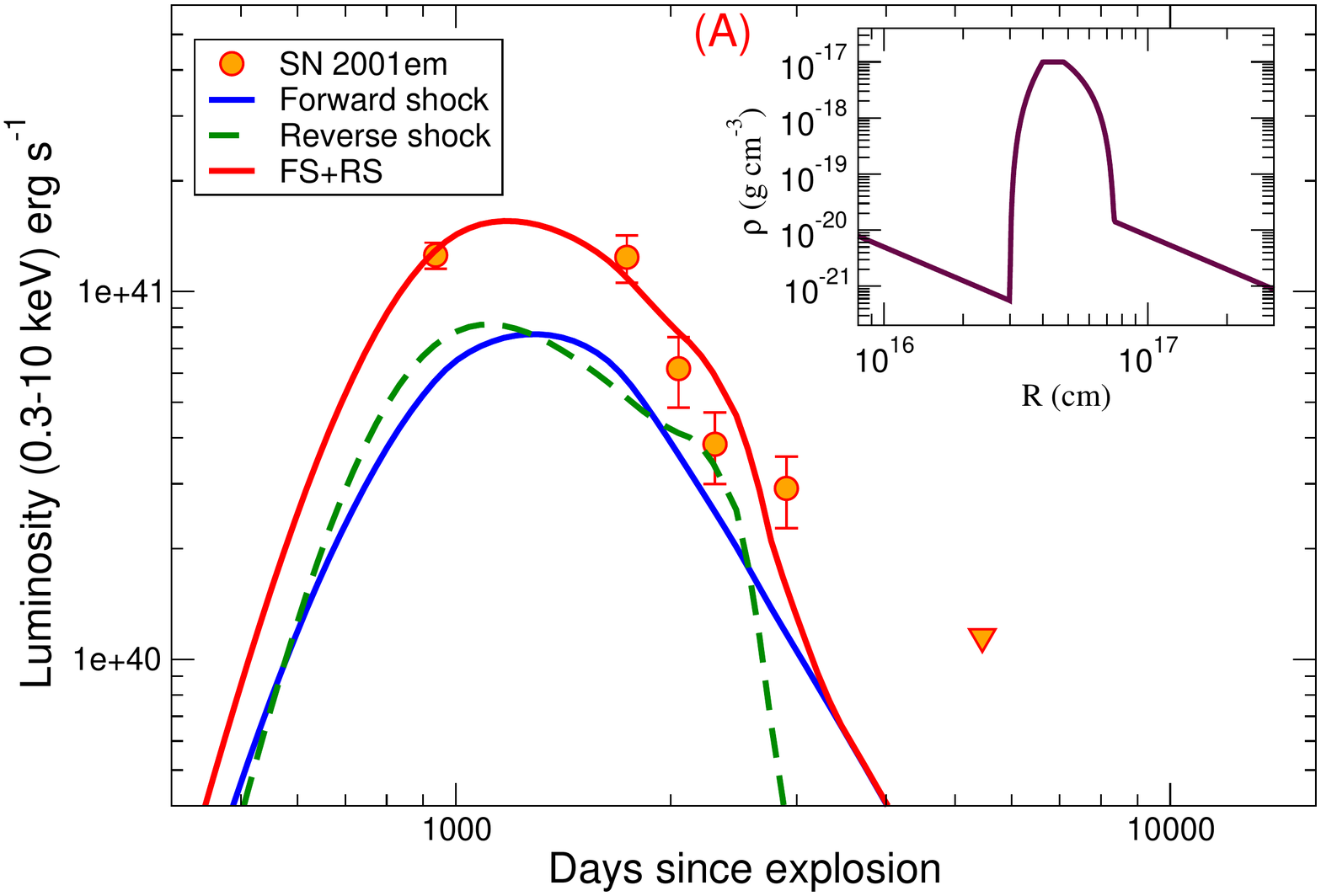}
\includegraphics*[width=0.49\textwidth]{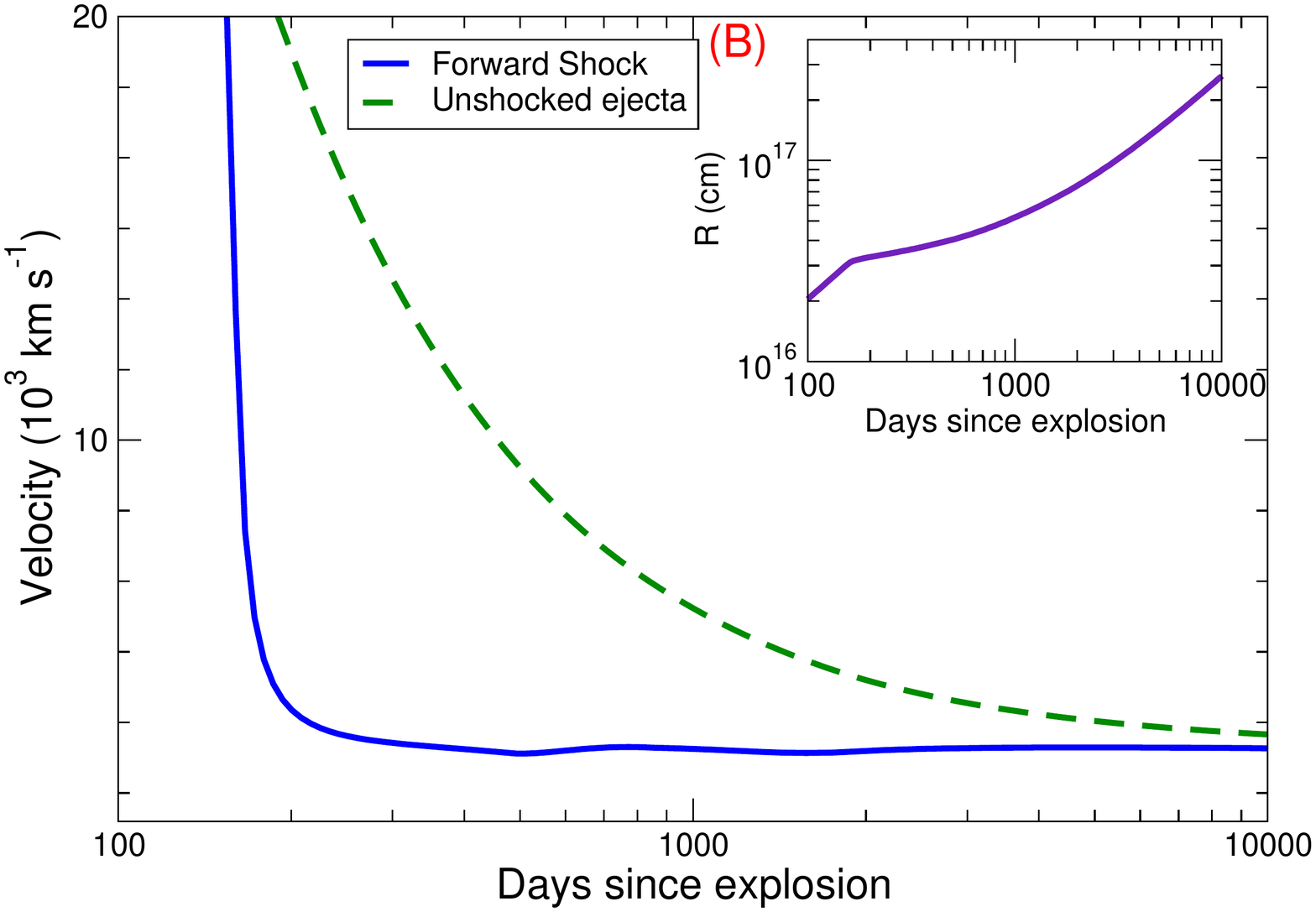}
\includegraphics*[width=0.49\textwidth]{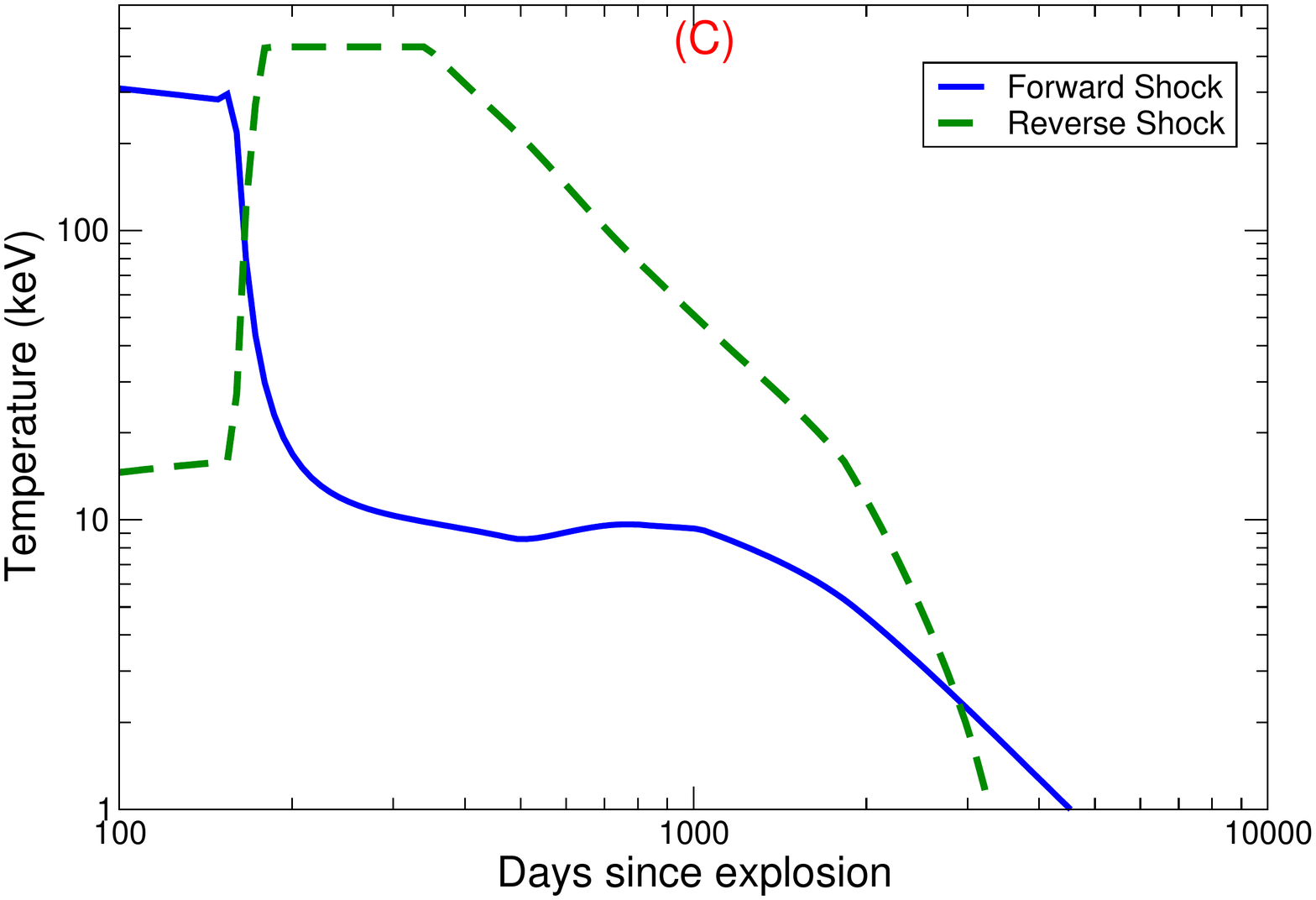}
\includegraphics*[width=0.49\textwidth]{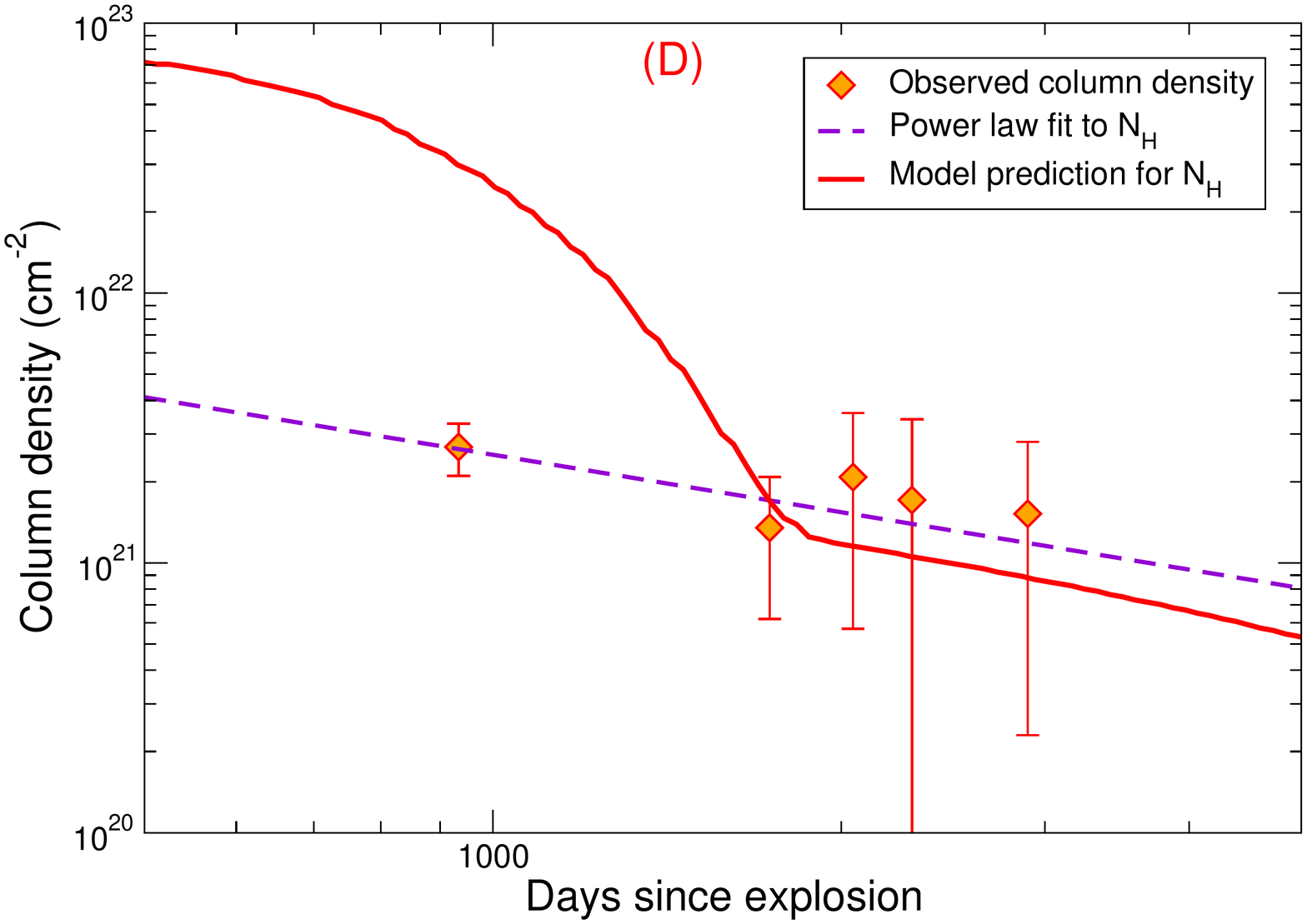}
\caption{
Evolution of X-ray luminosity, expansion velocities and shock temperature 
in the CS interaction model. Panel {\it A} shows the X-ray luminosity 
({\it red} line) composed of the emission from the forward shock ({\it blue solid} line) and reverse shock ({\it green dashed }) with the overplotted observational 
data;  the {\it triangle} symbol indicates an upper limit. The inset shows 
the CS density distribution. Panel {\it B} shows the velocity of the thin
dense shell ({\it  blue solid} line) and the maximal velocity of unshocked 
ejecta ({\it dotted}). The radius of the thin shell is shown in the inset.
Panel {\it C} shows the temperature in the forward shock ({\it  blue solid} line) 
and the reverse shock ({\it dotted}). Panel {\it D} shows the best fit X-ray column densities. The evolution is consistent with a power law index of  $-0.71\pm0.23$. Model fit to $N_H$ is for a 
fiducial value of the outer  density wind profile of  $\dot M/u_w=10^{15}$\,g\,cm$^{-2}$. }
\label{fig:xmod}
\end{figure}

We carried out detailed modelling of  the X-ray emission following the recipe of    \citet{cc06}, but with 
modest revision.
 The original \citet{cc06} model treats the CS interaction in the thin shell approximation.
 The density profile of the freely expanding ejecta is an inner flat region outside of which there is a steep power law, $r^{-9}$.
The thermal state of shocked gas in the forward and reverse shocks is described  
 taking into account adiabatic and radiative cooling in the energy equation with 
  volumes of post-shock gas specified by the relative width 
  $\Delta R/R = 0.2$ and 0.1  for the forward and reverse shocks,  respectively.
 The SN ejecta of the former model is set by a mass of  $2-3M_{\odot}$ and a
 kinetic energy  of $1.6\times10^{51}$\,erg. 
 The CSM is composed of the inner rarefied WR wind and 
 the outer RSG wind with the  massive dense CS shell ($M \sim 2-3M_{\odot}$) at the radius 
 of $(5-7)\times10^{16}$\,cm. 
The CS shell should be  clumpy in order to avoid strong absorption of 
  X-rays with an energy of $\lesssim 1$ keV from the reverse shock; the required 
  occultation optical depth is $0.5-1$ \citep{cc06}.

The moderately modified model we use here (Fig. \ref{fig:xmod}) is specified by  
   the ejecta mass of $3M_{\odot}$, energy of $10^{51}$\,erg, 
   the CS shell ($3M_{\odot}$) at the distance of $(4-5)\times10^{16}$\,cm and 
   the similar inner and outer winds.
   The model predicts a significant contribution from both forward and reverse shocks, and fits 
 the data 
  satisfactorily  (Fig. \ref{fig:xmod}A).
The model gas temperature of the reverse shock 
  on day 1735 is  20 keV (Fig. \ref{fig:xmod}C).
However, both reverse and forward shocks are adiabatic at $t < 2700$ d.  Later on the 
  reverse shock becomes radiative, which is marked by the rapid temperature fall 
  (Fig. \ref{fig:xmod}C). The adiabaticity is in contrast to the analytical analysis and modeling of the radio data.
  The predicted column density at early epochs $\le 1600 $\, days is  higher than the observed one  (Fig. \ref{fig:xmod}D).
  However, this  issue may be resolved taking into account asymmetry and/or the clumpy structure of the CS shell, in which 
  case the H$\alpha$ emission  originates from the shocked clumps with  efficient cooling.

Summing up, in the modified  \citet{cc06}  model, the X-ray emission 
  arises from the contribution of both reverse and forward shocks driven by the ejecta 
  interaction with the massive clumpy CS shell, while the H$\alpha$ emission is related 
  to the shocked clumps of this CS shell. 
  However, this model should be considered as a
simplified description of the three-dimensional (3-D)
situation. In the scenario  of a clumpy shell the
H$\alpha$ width is related to cloud shocks, whereas
the speed of the interclump forward shock should
significantly
exceed 2000 \kms,  and generally be consistent with the 4000
\kms\
supported by other observational constraints.
  A better handle on multi-waveband data and a  more realistic model 
with  3D-hydrodynamics with a proper treatment 
  of thermal and radiation processes in a multi-phase medium are needed, but are  out of the scope of this paper.

 \section{Discussion} \label{sec:discussion}
 
 \subsection{Phases of circumstellar interaction in SN 2001em}

  We have carried out a comprehensive analysis of radio and X-ray emission from \snem. 
  We have qualitatively defined the evolution in three phases. 
 We did not catch the interaction in  Phase - 1  because by the time the  radio and X-ray observations commenced, the supernova 
 ejecta were already interacting with the shell. 
  Our model favours a forward shock
  origin for the X-ray emission. The modified  \citet{cc06} model
suggests X-ray emission having a contribution from both shocks, but requires both shocks to be adiabatic for up to 2700 days and overpredicts the
early column density.   These issues can be reconciled if one assumes a clumpy CSM shell.  
We note that  asymmetry and clumping are important, so it is difficult to  fit a convincing model for both the X-ray and radio emission.

\begin{figure*}
 \centering
\includegraphics*[angle=0, width=0.65\textwidth]{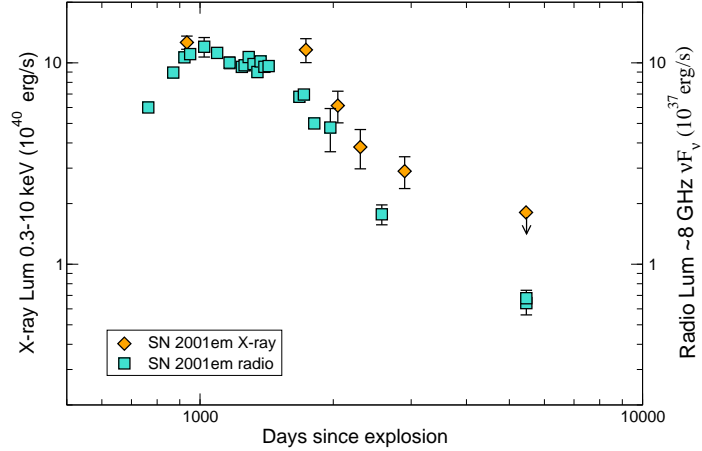}
\caption{The $\nu \rm F_\nu$ plot for \snem\ X-ray (0.3--10\,keV) and radio data  at 8.5 GHz. The left side y-axis is labelled for 
the X-ray flux and the right side y-axis is labelled for the radio flux.
\label{fig:nuLnu}}
\end{figure*}

 \citet{cc06} in their original analysis
 had considered two models: one where the shock had overtaken the dense shell by the time X-ray observations commenced and another where the shock was still
 in the dense shell. Since the data at that time covered only the first 1000 days, they could not distinguish between the two models. 
Our fits to radio observations suggest evolution in Phase -- 2 but  do not
show when  Phase -- 2 ended. 
The indications for this come from the X-ray observations. 
The X-ray light curves start to decline faster after the \xmm\ epoch (i.e. 1735 days), suggesting that Phase-3 already started at this epoch (Fig. \ref{fig:nuLnu}). 
The fast decline suggests  a change in the ejecta-CSM interaction evolution. 
We suggest that the SN ejecta interacted with the dense shell until or before the \xmm\ epoch and then the shock came out of
 the  shell and started running through a faster declining CSM.  
  
\subsection{Masses in the cool dense shell and the CSM H-shell}

We  estimate the masses in various components of shocks and shells.
The radio data are best fit by a combination of internal and external FFA, with a dominant contribution from internal FFA. Internal FFA is possible if
 a fraction of the cold gas is mixed in
the radio emitting region.  This is similar to what seen in SN 1986J where the SN radio
light curves were best fit with the internal absorption model mixed with external absorbers  \citet{weiler+90}.  
The fact that the spectral index evolution is similar in both SNe (Fig. \ref{fig:alpha86j})  
suggests that similar mechanisms determine the light curves.
Similar results have been found  for other SNe IIn, e.g.,   SN 1988Z \citep{vandyk+93, williams+02}
and  SN 2006jd \citep{chandra+12b}. 
In SN 2006jd, \citet{chandra+12b} calculated that  a modest amount of cool gas mixed into the emitting region 
was sufficient to give rise to the needed absorption.

In the case of SN 2001em, we can do a rough estimate of the amount of mixing of the thermal absorber in
the synchrotron emitting region if we assume the absorbing gas is in pressure equilibrium with the X-ray
emitting gas and that it has a relatively low temperature ($10^4-10^5$ K). 
We calculate the mass loss when there is an optical depth of unity  at 8.46 GHz, at an age of $\sim
 960$ days. The mass-loss rate at this epoch is $ \sim 2\times10^{-3}$\,\mdot, corresponding to a circumstellar density of $9.3\times 10^5$ cm$^{-3}$.
The X-ray producing shock has an estimated  temperature of  $\sim20$ keV, corresponding to 
 a shock speed of $v_{s}\sim  4000$ \kms. 
Following the analogy of SN 2006jd  \citep{chandra+12b}, a small amount of mixing of cool gas, i.e. $\le10^{-8} \,M_\odot$
can give rise to the needed absorption.
The source of the cool gas is likely to be radiative cooling of dense gas in the shocked
region, suggesting that the cool shell  likely  formed between the forward and the  reverse shock.

We can also  estimate the mass and energetics of the dense CSM shell.
The observed shell interaction occurred roughly   between days 764 and 1735. If we assume that the shock entered the dense shell not long ago before the first observations, then the duration of interaction is  around 1000 days. For a velocity of the forward shock of 4000\,\kms, the corresponding  shell width is $3.5 \times10^{16}$\,cm.
Our estimate of the density  $9.3\times 10^5$ cm$^{-3}$ implies  the mass of the shocked CSM in the shell is $M_{\rm shell} \sim 2.5\,\rm M_\odot$.
The total kinetic energy of the swept up material is $\sim E_{\rm shell}=0.5 M_{\rm shell}  v_{\rm FS}^2 \sim 4\times 10^{50}$ erg.

 \subsection{Late time spectrum of SN 2001em}

  \begin{figure}
  \centering
\includegraphics*[angle=0, width=0.8\textwidth]{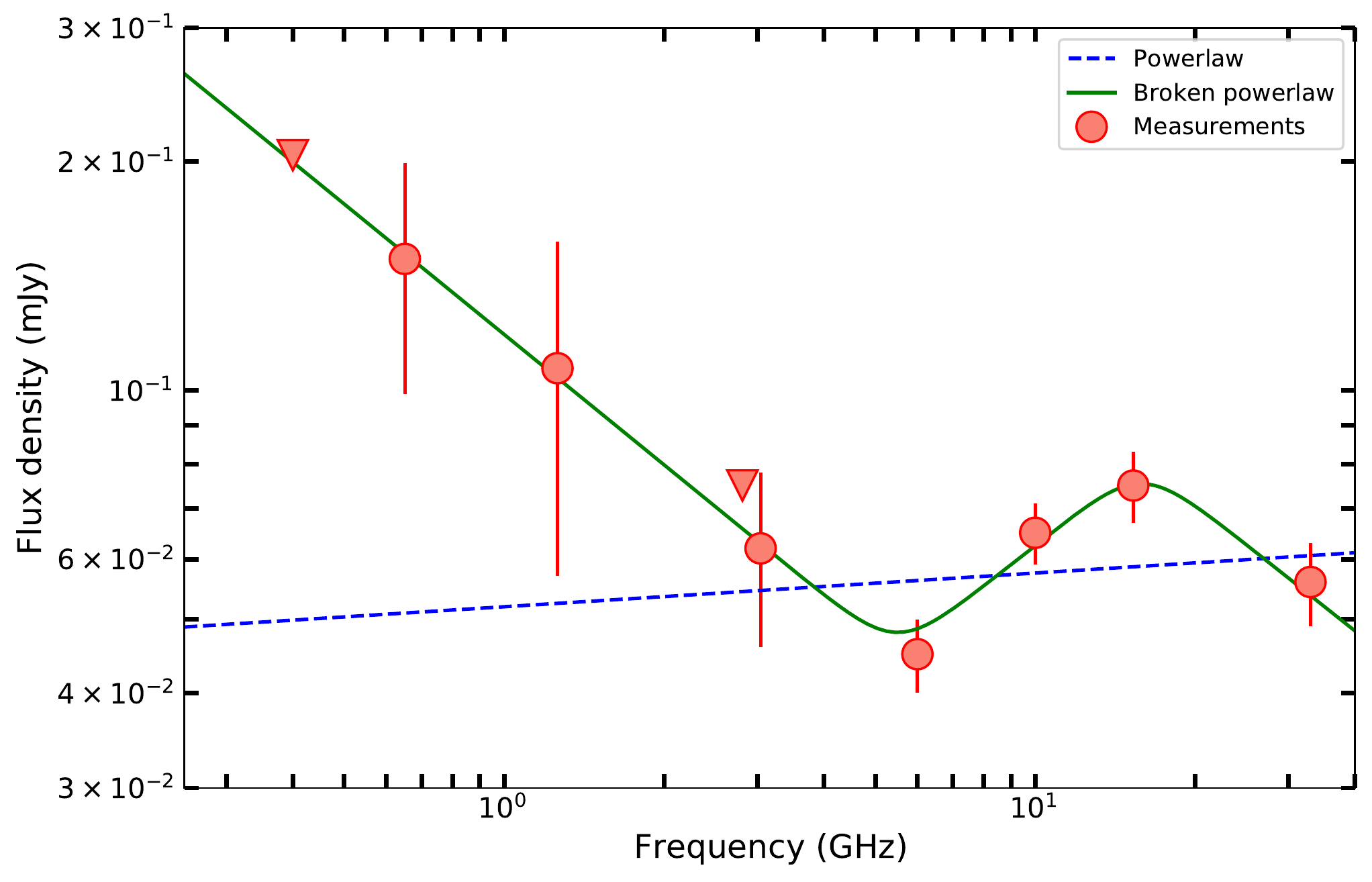}
\caption{Day 6800 radio spectrum of SN 2001em. We carried out an MCMC analysis to fit a simple powerlaw and a double broken powerlaw model.
 \label{fig:day6800}}
\end{figure}

 We have late time data at around day 5460, day 6670 and day 6872. We combine the last two epochs to create a wide band radio  spectrum
 at an average epoch of day 6772..
 The standard  radio emission model for supernovae predicts optically thin spectra at these late stages. 
 At  day 5460, the spectra cover the frequency range 3--11 GHz and the spectrum is optically thin.  On day 6772, the 
 spectrum covers a frequency range of 0.4--32 GHz. Visually, the spectra are optically thin up to around 6 GHz, become optically thick,
 and then have a second transition to an optically thin phase  at around 15 GHz. To estimate the robustness of these two transitions, we
 carry out 
Markov Chain Monte Carlo (MCMC) fitting  using the Python package emcee \citep{fm+13}. 
We choose 200 walkers, 1000 steps, and flat priors on all of the parameters.
We obtain best-fit values
(at 68\% confidence interval, i.e., $1\sigma$). We fit a smoothed broken powerlaw model with two break frequencies, as well as a single powerlaw
model (Fig. \ref{fig:day6800}). A single powerlaw model results in an index of $0.04^{+0.08}_{-0.08}$ with a reduced $\chi^2=3.3 $.
The broken powerlaw results in indices between transition frequencies to be 
$-0.57^{+0.08}_{-0.08}$, $+0.56^{+0.03}_{-0.04}$ and  $-0.55^{+0.07}_{-0.09}$, respectively with a reduced $\chi^2=0.89 $.
We show the fits in Fig. \ref{fig:day6800}, indicating that the significance of the transitions seen in the spectrum is quite robust.

\begin{figure}
  \centering
\includegraphics*[angle=0, width=0.8\textwidth]{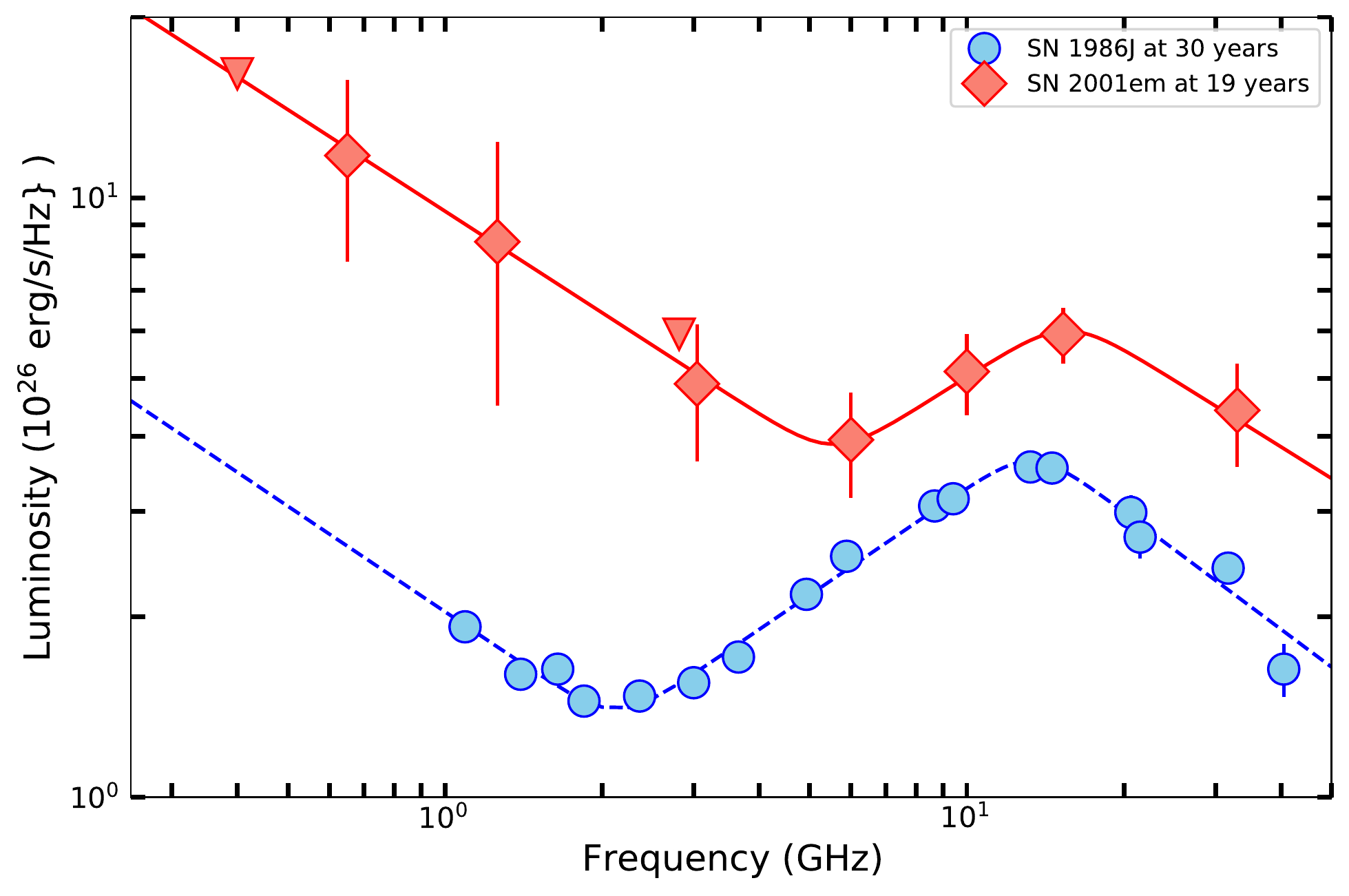}
\caption{Comparison of the late time radio spectrum of SN 2001em at  $\sim 19$ yrs with that of SN 1986J at $\sim 30$ yrs, which showed  the presence of a central component at late times \citep{bietenholz+17}.
\label{fig:86j}}
\end{figure}

The late time spectrum in \snem\, is quite similar to that of SN 1986J (Fig. \ref{fig:86j}).
In SN 1986J, the spectrum  at low frequencies had the form of
a simple power law typically seen for SNe, with spectral index, $\alpha \approx -0.5$, but the radio emission was partly absorbed
at high frequencies.
Late VLBI observations showed a dominant central component  \citep{bietenholz+17}. 
In this scenario the best 
interpretation for \snem\ is that the low frequency optically thin emission is from the fast shocks, whereas the rising spectrum at higher frequencies unveils the 
hidden central absorbed component as in SN 1986J.

\subsection{A common envelope scenario?}

In general, some supernova remnants have been shown to expand in W-R bubbles, before hitting the dense shell created by slow red-supergiant
winds. However, in normal cases, the W-R bubble radius is around 10--12 pc, and the SN shock takes around 800--900 years to reach the shell \citep{d05}.
In SN 2001em, and a few other known cases, the shock hits the shell in a few hundred days, i.e. the outer envelope was lost only a few 
thousand years
before the explosion.  How do these supernovae lose their  envelope so close to explosion is not clearly understood.
 Our mass-loss rate estimate, $\sim (1-5)\times 10^{-3}$\,M$_\odot$\,yr$^{-1}$, is   high for such late time interaction.  
The supernova lost a  $\sim 2.5$\,M$_\odot$ H-envelope.

This is a very large mass to
be lost and the mechanism for the mass loss is not clear.   
Such large mass loss does occur during giant eruptions of Luminous Blue Variables (LBVs); $\eta$ Carina is an example.
Interestingly,  a binary merger has been suggested for $\eta$ Carina, in which there is a dense disk in the orbital plane and faster, lower density regions away from the plane \citep{smith+18}
However, the LBV phase is generally not expected to be close in time to a supernova.
A binary scenario involving a neutron star is a possible
option, in which an in-spiraling neutron star causes mass loss followed by an explosion when it gets to the center of the star
\citep{chevalier12}.
We suggest below  that this
 could be a plausible scenario, support for which comes from the late time radio observations.  

The explanation of the unusual spectrum in SN 1986J is that the radio  was dominated by the  standard optically thin  expanding shell  at low frequencies.
The partially absorbed high frequency emission was dominated by the central component seen  in VLBI observations, which eventually became optically thin at even higher frequencies.
This behaviour might be explained by a  common-envelope scenario in which  a highly structured CSM is produced by a binary companion, as suggested by \citet{chevalier12}. In 
this scenario, the ejecta are initially optically thick, and
 the ejecta expanding above and below the equatorial disk
 initially absorb  emission arising in these regions. After a time delay  the central
disk can be seen and the radio central component
starts to reveal itself. 
In SN 1986J 
 the central component was explained to be due to the
SN shock interacting with the highly structured CSM
produced by a binary companion, where the
shock travelled much more slowly in the denser parts of the CSM
near the binary orbital plane, thus producing a bright but compact
radio emission region \citep{bietenholz+17}. A similar explanation may hold  for \snem.
To investigate the relation between SN 1986J and \snem, we
 overplot the spectrum of SN 1986J at 30 years with that of SN 2001em on day 5400 (Fig. \ref{fig:86j}). 
The similarity between the spectra in the overlapping frequency range is remarkable. 

The common envelope scenario  can also  explain the loss of a 2.5\,M$_\odot$ envelope, as the 
high  mass loss is driven by common envelope evolution of a compact object in the envelope of a massive star and the SN explosion  is triggered by the inspiral of the compact object to the central core of the companion star.  

A prediction of a common envelope scenario is an  asymmetric CSM.  Our X-ray data provide evidence of asymmetry.  Continuing radio observations over a wide band are needed to follow the spectrum of \snem\  at late epochs to see whether there is evolution toward optically thin emission as is observed in SN 1986J.

\subsection{Comparison with  normal SNe IIn}

\begin{figure*}
\centering
\includegraphics*[angle=0, width=0.79\textwidth]{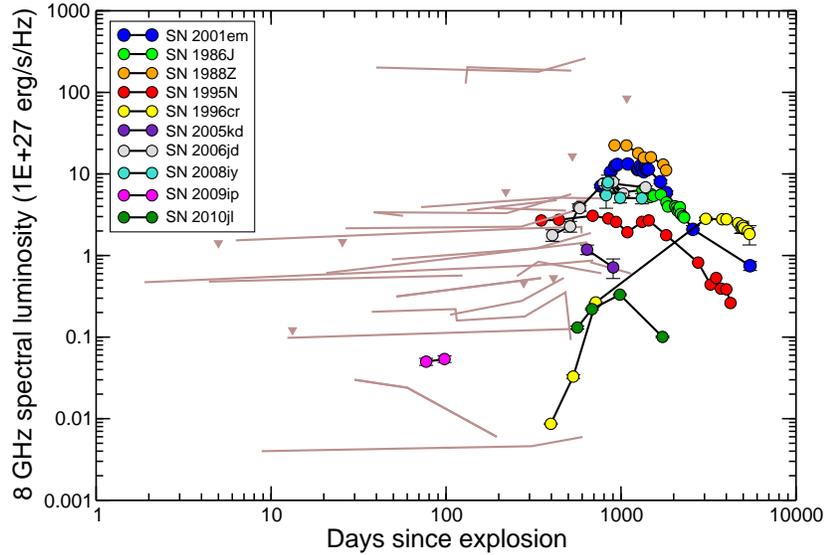}
\caption{Comparison of the \snem\ 8 GHz radio spectral luminosity to some of the well observed SNe IIn. Circles are detections; the lines indicate upper limits for observations at many epochs. Single triangles indicate upper limits at individual epochs (supernovae that were observed only once). \label{fig:01em-radio}}
\end{figure*}

In Fig. \ref{fig:01em-radio}, we compare the SN 2001em 8 GHz spectral luminosity light curve with  those of published SNe IIn light curves.  \snem\ is the second brightest radio supernova among the 
class of observed SNe IIn. 
Radio emission in SNe IIn is quite intriguing and most of the SNe IIn seem to become detectable at late epochs, after a few hundred days. This can partly be attributed to 
observational bias as some of the 
 SNe IIn were either discovered late or observed late. For example, the  prototypical SNe IIn SN 1986J, SN 1988Z and SN 1978K were discovered/observed years after their explosions. 
 Many of the SNe IIn discovered sooner were not observed early enough in radio bands. 
 However, SN 2010jl radio observations started by day 45, but the first radio detection was reported after 500 days, which \citet{chandra+15} attributed to efficient absorption of radio emission at early epochs.
SN 2009ip is the only exception.  It was detected in radio bands early on but faded below detection within a few tens of days. However, SN 2009ip was a peculiar explosive event  \citep{margutti+14}.  

Thus late radio turn on in SNe IIn, with the exception of SN 2009ip,  is a  common phenomenon. It remains to be seen in how many cases it can be  attributed to high density absorption, and in  how many cases due to  late discovery/classification.
For SN 2001em, the frequency dependence of the radio evolution indicates that absorption plays a role in the evolution.

\begin{figure*}
\centering
\includegraphics*[angle=0, width=0.79\textwidth]{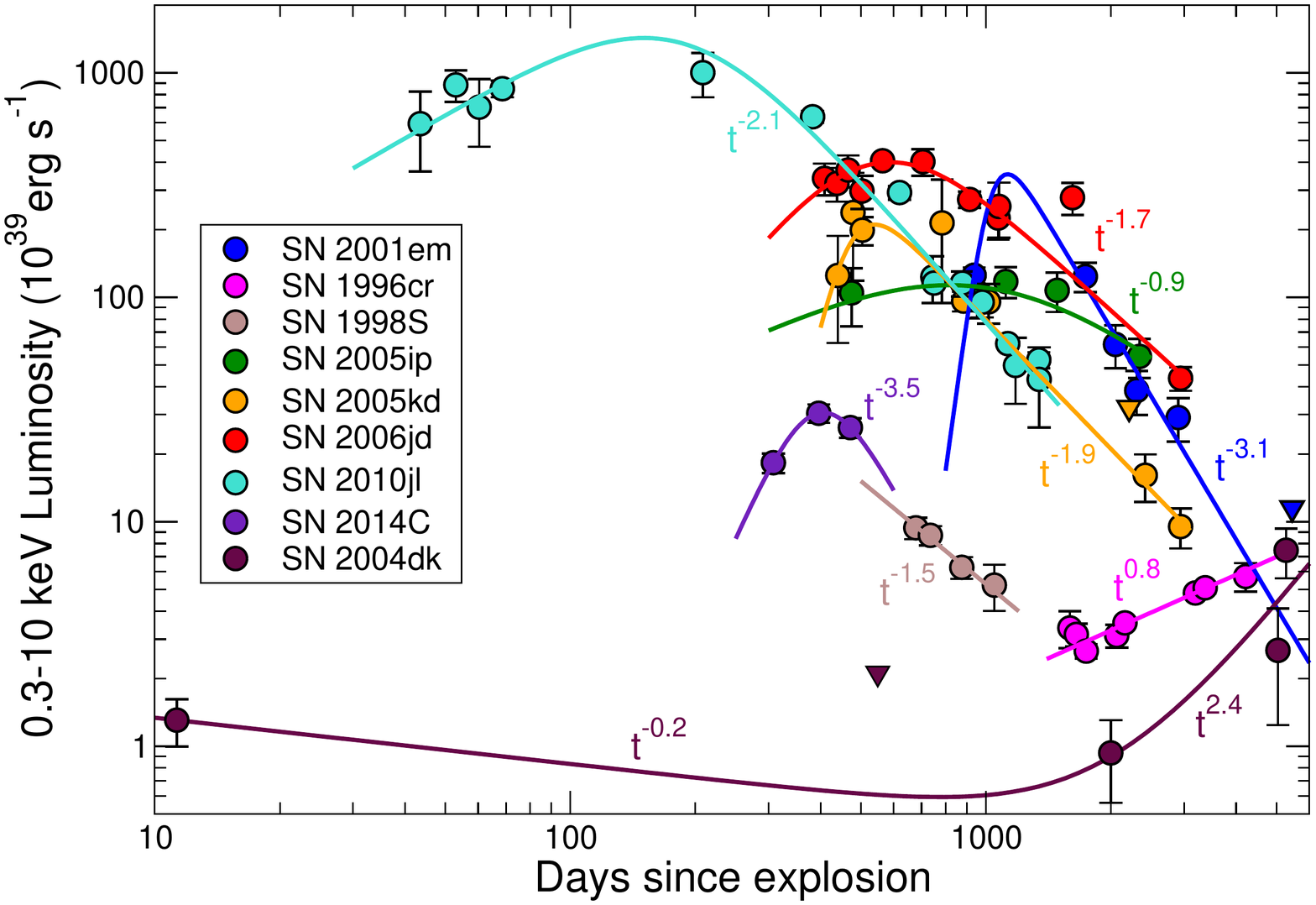}
\includegraphics*[angle=0, width=0.79\textwidth]{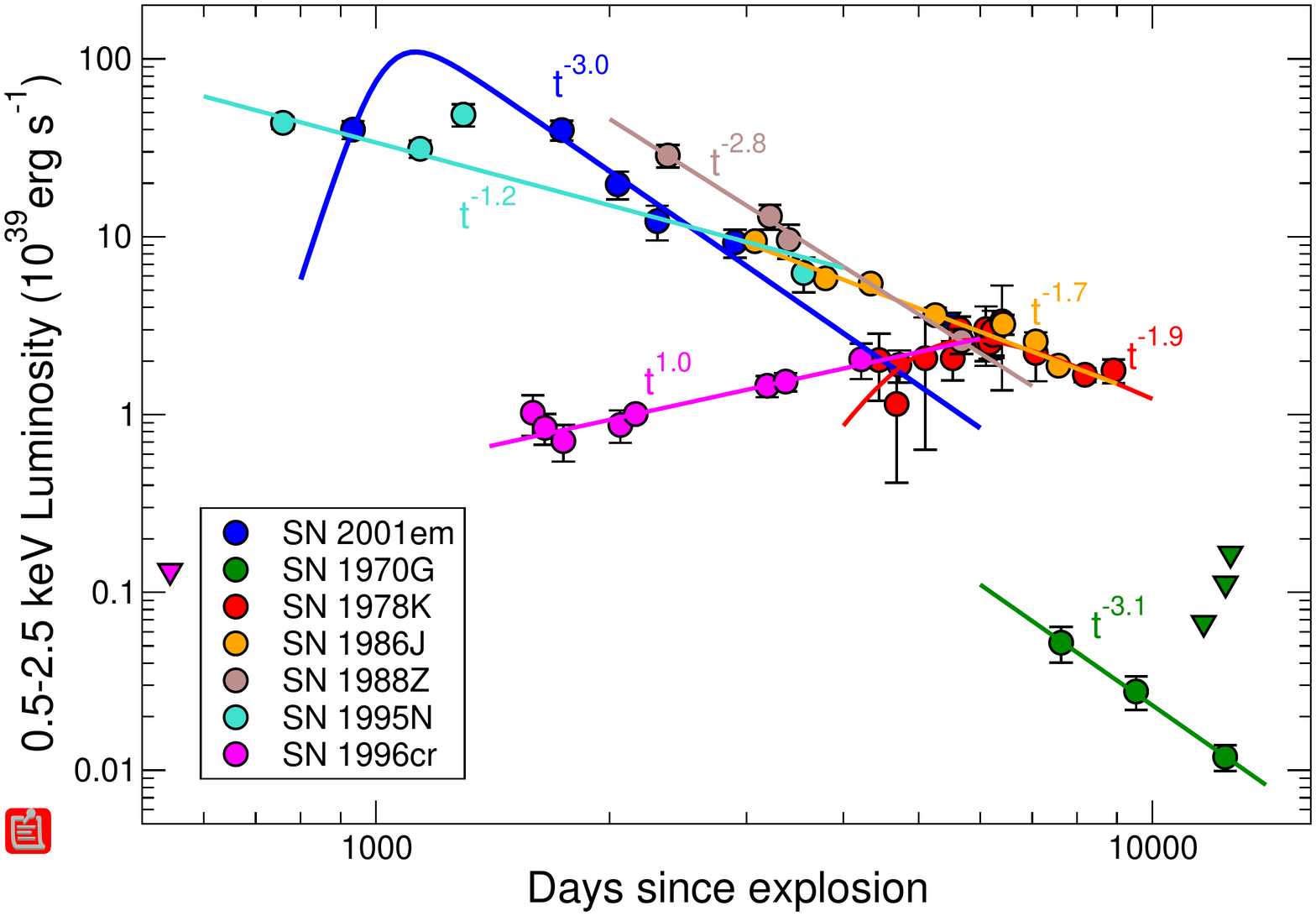}
\caption{Comparison of the \snem\ X-ray light curve with several other well observed light curves of interacting SNe IIn. The top panel shows the 
SNe IIn which have measured fluxes in the 0.2--10\,keV energy range. 
The bottom panel shows SNe IIn in the soft energy band, which are mainly old
SNe IIn observed with  ROSAT. For comparison, we have estimated the \snem\ flux in the soft energy band by using the fit parameters in 
Table \ref{tab:xrayobs}.  Many supernovae deviate from the standard $1/t$ dependence.
The early evolution of SN 2001em is poorly constrained.   \label{fig:01em-xray}}
\end{figure*}

The forward shock X-ray luminosity of a supernova, in an ideal case with no cooling and bremsstrahlung emission  in a steady wind environment,  is expected to decrease as $t^{-1}$. 
We observe a faster decline in the X-ray flux at the late stages of \snem\ and interpret
 this as the shock moving into a medium with steeper density decline than $s=2$. 
  \citet{dwarkadas+12} found that the  evolution expected for interaction with an $s=2$ steady wind is rarely seen in 
  supernovae, especially SNe IIn. 

In Fig. \ref{fig:01em-xray}, we plot 0.2--10\,keV X-ray light curves of various SNe IIn.
We also compared the behaviour of SN 2001em with older SNe IIn, which had most of the observations in  the
{\rm ROSAT} soft X-ray bands. For this comparison, we obtained the SN 2001em flux in the 
0.5--2.5\,keV energy range. SN 2010jl, SN 2006jd and SN 2005kd also show 
 fast declining evolution. 
SN 2010jl  showed a  flat evolution for the first 300 days followed by a more rapid decline.
 SN 1988Z and SN 1970G also show very steep X-ray evolution. The luminosity evolution of SN 1988Z shows a decline slope of 
$-2.6\pm0.6$ \citep{schlegel+06}. In SN 1970G,
the  slope was $-2.7\pm0.9$ \citep{sk+05}.
SN 1986J was caught much later but it also revealed a steep
luminosity decline \citep{houck+05}. 
Since the shock is expected to undergo adiabatic cooling and the temperature is expected to move to lower values, one 
does not expect a steepening at late times due to spectral constraints of the telescopes. The most plausible mechanism is that the
shock is moving into a  lower density medium with $s>2$.
In  SN 2010jl, \citet{chandra+15} also  attributed the faster decline  to a more rapidly decreasing circumstellar density profile. However,
in some cases different parameters could give a situation in which the reverse shock wave has moved in to the flat part of the ejecta density distribution, which would also give a more rapid decline \citep{ofek+14}.

\subsection{Comparison with supernovae undergoing metamorphosis}

\snem\ is an example of a class of core-collapse supernovae that fill a gap between events that interact strongly with nearby environments immediately after explosion (Type IIn and Ibn) and events that are not 
observed to interact. In \snem, the interaction started after a few hundred days when the shock caught up to the ejected outer H layers of the pre-SN progenitor star.  The details of shedding the outer
layers in massive stars are not well known, though the common-envelope scenario is a possible mechanism in \snem.

\begin{figure*}
\centering
\includegraphics*[angle=0, width=0.75\textwidth]{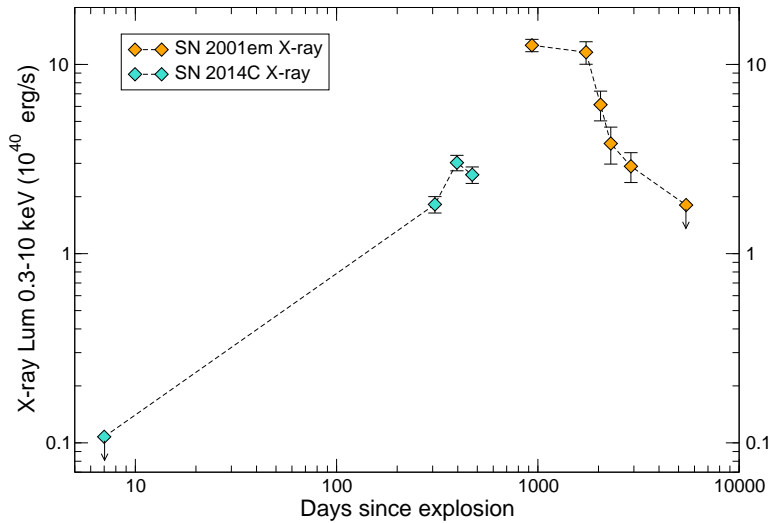}
\caption{X-ray evolution of SN 2001em compared with that of SN 2014C \citep{margutti+17}. 
 With its early data, SN 2014C provides a complementary point of comparison to SN 2001em.  \label{fig:with14c}}
\end{figure*}

SN 2014C is a younger counterpart of \snem\  that   also had a stripped-envelope progenitor, and  started showing  similar signs of
interaction by day 100   \citep{milisavljevic+15}.  \citet{margutti+17} found that in the case of SN 2014C, the 1\,$M_\odot$ H-shell was ejected only decades to centuries before the SN explosion.
  \snem\ has revealed slower evolution, and the mass-loss episode leading to the shell happened $200 - 400$ years prior to the explosion. 
  Possible reasons are that the shell was ejected much earlier in \snem\
 or that the progenitor winds were faster and
 accelerated the shell more rapidly. 
The early observations of SN 2014C complement SN 2001em and may give a qualitative picture of these SNe at early stages (Fig. \ref{fig:with14c}).

SN 2004dk is another case of metamorphosis. 
SN 2004dk was a stripped envelope supernova with very detailed radio observations that revealed late time 
rebrightening \citep{wellons+12}. Its X-ray luminosity around 13 days was $2\times 10^{39}$ erg s$^{-1}$ \citep{wellons+12}.
Deep late-time optical spectroscopy at  4684 days showed H$\alpha$ features, revealing that the shock  caught up with the
lost H \citep{mauerhan+18}.  \citet{pooley+19} presented observations spanning 15 years in X-rays. They estimated
that the shell of H-rich wind material was ejected $\sim 1400$ years prior to the explosion,  during carbon burning.

SN 2017ens was also a broad-line  Type Ic supernova that changed into a Type IIn supernova
with 2000 \kms\  H$\alpha$ emission with a high luminosity $L_{\rm H\alpha}=3\times10^{40}$\,\ergs
\citep{chen+18}.  Another example is SN 2005bf which was an unusual stripped envelope supernova \citep{folatelli+06}: the explosion of 
a massive He star with a trace of a hydrogen envelope \citep{anupama+05}. A  Keck spectrum taken $\sim 233$ days after explosion showed strong, broad H$\alpha$ emission  similar to  SN 2001em. \citet{vandyk10}
suggested SN 2005bf and SN 2001em to be cases of a very massive star that experienced a powerful LBV eruption before evolving to the WR phase prior to explosion.
Peculiar supernova SN 1996cr was initially 
identified  as an ultraluminous X-ray source known as Circinus  X-2, but later classified as 
a core collapse SN of Type IIn \citep{bauer+08}.
\citet{dwarkadas+10} carried out detailed modeling and interpreted   SN 1996cr as exploding in a low-density medium before interacting with a dense CSM shell, caused 
by the interaction of a blue supergiant or Wolf-Rayet (W-R) wind with a previously existing red supergiant (RSG) wind, 0.03 pc away from the progenitor star. The data also revealed that the shock wave has now exited 
the shell and is expanding into a medium that is consistent with the  undisturbed continuation of the dense RSG wind. Another case is SN 2012ca. While initially it was thought to be a thermonuclear supernova, 
Public ESO Spectroscopy Survey of Transient Objects (PESSTO) nebular spectroscopy suggested it to 
be a stripped-envelope core collapse event, showing signs of interaction later on \citep{inserra+14}. 
SN 2012ca was marginally detected in X-rays  \citep{bochenek+18}.

A radio survey of stripped envelope SNe at late time carried out by \citet{bietenholz14} revealed that these  SNe  are few, yet the
handful of  examples pose questions for  current theories of massive star evolution. 
\citet{vinko+17} carried out a long-term survey  of hydrogen-poor supernovae, with the aim 
of discovering late-time interaction and  finding the hydrogen envelope expelled from the progenitor star several decades/centuries before the explosion. 
They detected continuum subtracted H$\alpha$ emission in 9 SN Ibc, 1 SN IIb, and 3 SN Ia events. 
Similarly \citet{margutti+17} found  luminous radio rebrightening in SN 2003gk, SN 2007bg, and SN iPTF11qcj, attributing it to
 the ejecta-CSM collision. 
These examples  indicate that there is probably a continuum between stripped envelope supernovae to 
hydrogen rich interacting supernovae.

\subsection{Similarities with SN 1986J}

SN 1986J deserves special attention due to it's similarities with \snem. It was discovered a few years after the explosion \citep{rupen+87}, so it is unclear whether it was a Type IIn supernova or a 
stripped-envelope supernova which underwent metamorphosis. 
 Both supernovae are best fit with a $s=2$ medium and the  internal absorption is dominant in both cases \citep{weiler+90}. 
Fig. \ref{fig:alpha86j} shows the evolution of spectral indices in the two supernovae, which is strikingly similar.

In the other younger cousin, SN 2014C, the ejecta speed was much higher than that of SN 2001em; the expansion speeds were  similar in \snem\ and SN 1986J.
The VLBI measurements for \snem\ resulted in an expansion speed of $< 6000$ \kms \citep{schinzel+08, bietenholz+05}, SN 1986J was 
found to have an expansion speed at around 1000 days to be $5700$\,\kms \citep{bietenholz+04}. 

The most striking similarity is in the  radio spectra at late epochs (Fig.  \ref{fig:86j}). 
VLBI observations have ascertained that in SN 1986J, this optically thick spectrum is arising from a different interior component which is unveiled at a later time due to decreasing absorption   \citep{bietenholz+17}. A common-envelope scenario is a possible explanation for both supernovae \citep{chevalier12}. Thus \snem\ is most likely the younger cousin of SN 1986J.

\section{Conclusions}
\label{sec:conclusion}

\snem\ is the  oldest known SN to undergo metamorphosis from non-interacting to interacting supernova.  Its long term study has provided us a unique opportunity to probe its pre-explosion
history. 
Comprehensive analysis of radio and X-ray data indicate that the   \snem\ ejecta entered the shell  before the first radio observations
 around 760 days, i.e. the end of Phase-1.
The SN ejecta subsequently shocked the dense shell
for around 1000 days.
After around 1700 days the forward shock wave came out of the shell and started moving in the fast declining wind. Signatures of this behavior are seen in the X-ray observations.  Our modelling suggests that radio and X-ray emission are primarily arising from   the forward shock. Both radio and 
X-ray measurements reveal asymmetry and clumpiness in the ejecta and the CSM medium.
One of the most interesting features is the  optically thick radio  spectrum after around 5000 days, which  can be interpreted to be coming from a central component, 
similar to that seen in SN 1986J.
Common envelope evolution  may have played a role in producing the dense surroundings.
The late time observations have given a new twist in the story of \snem.
We will be able to probe it further  with future observations.

\acknowledgements
We thank referee for very useful comments which improved the manuscript significantly.
PC acknowledges the support from the Department of Science and Technology     via
SwarnaJayanti Fellowship awards (DST/SJF/PSA-01/2014-15).
RAC acknowledges support   from   NSF grant     AST-1814910 and from
NASA through Chandra
award number AR8-19007X issued by the Chandra X-ray Center.
The Chandra Center is operated by the Smithsonian Astrophysical Observatory
for and on behalf of NASA under contract NAS8-03060.
 The National Radio Astronomy Observatory is a facility of the National Science Foundation operated under cooperative agreement by Associated Universities, Inc.
This research has made use of data obtained through the High Energy Astrophysics Science Archive Research Center Online Service, provided by the NASA/Goddard Space Flight Center.
We thank the staff of the GMRT that made these observations possible.
The GMRT is run by the National Centre for Radio Astrophysics of the Tata Institute
of Fundamental Research.
This work made use of data supplied by the UK Swift Science Data Centre at the University of Leicester.

\vspace{5mm}
\facilities{CXO, Swift(XRT), VLA, GMRT,
XMM}

\bibliography{ms}{}

\begin{thebibliography}{}
\expandafter\ifx\csname natexlab\endcsname\relax\def\natexlab#1{#1}\fi

\bibitem[{{Anupama} {et~al.}(2005){Anupama}, {Sahu}, {Deng}, {Nomoto},
  {Tominaga}, {Tanaka}, {Mazzali}, \& {Prabhu}}]{anupama+05}
{Anupama}, G.~C., {Sahu}, D.~K., {Deng}, J., {et~al.} 2005, \apjl, 631, L125

\bibitem[{{Arnaud}(1996)}]{arnaud+96}
{Arnaud}, K.~A. 1996, in Astronomical Society of the Pacific Conference Series,
  Vol. 101, Astronomical Data Analysis Software and Systems V, ed. G.~H.
  {Jacoby} \& J.~{Barnes}, 17

\bibitem[{{Bauer} {et~al.}(2008){Bauer}, {Dwarkadas}, {Brandt}, {Immler},
  {Smartt}, {Bartel}, \& {Bietenholz}}]{bauer+08}
{Bauer}, F.~E., {Dwarkadas}, V.~V., {Brandt}, W.~N., {et~al.} 2008, \apj, 688,
  1210

\bibitem[{{Bietenholz}(2014)}]{bietenholz14}
{Bietenholz}, M.~F. 2014, \pasa, 31, e002

\bibitem[{{Bietenholz} \& {Bartel}(2005)}]{bietenholz+05}
{Bietenholz}, M.~F., \& {Bartel}, N. 2005, \apjl, 625, L99

\bibitem[{{Bietenholz} \& {Bartel}(2007)}]{bietenholz+07}
---. 2007, \apjl, 665, L47

\bibitem[{{Bietenholz} \& {Bartel}(2017)}]{bietenholz+17}
---. 2017, \apj, 851, 7

\bibitem[{{Bietenholz} {et~al.}(2004){Bietenholz}, {Bartel}, \&
  {Rupen}}]{bietenholz+04}
{Bietenholz}, M.~F., {Bartel}, N., \& {Rupen}, M.~P. 2004, Science, 304, 1947

\bibitem[{{Bochenek} {et~al.}(2018){Bochenek}, {Dwarkadas}, {Silverman}, {Fox},
  {Chevalier}, {Smith}, \& {Filippenko}}]{bochenek+18}
{Bochenek}, C.~D., {Dwarkadas}, V.~V., {Silverman}, J.~M., {et~al.} 2018,
  \mnras, 473, 336

\bibitem[{{Cash}(1979)}]{cash79}
{Cash}, W. 1979, \apj, 228, 939

\bibitem[{{Chandra}(2018)}]{chandra18}
{Chandra}, P. 2018, \ssr, 214, 27

\bibitem[{{Chandra} {et~al.}(2012){Chandra}, {Chevalier}, {Chugai}, {Fransson},
  {Irwin}, {Soderberg}, {Chakraborti}, \& {Immler}}]{chandra+12b}
{Chandra}, P., {Chevalier}, R.~A., {Chugai}, N., {et~al.} 2012, \apj, 755, 110

\bibitem[{{Chandra} {et~al.}(2015){Chandra}, {Chevalier}, {Chugai}, {Fransson},
  \& {Soderberg}}]{chandra+15}
{Chandra}, P., {Chevalier}, R.~A., {Chugai}, N., {Fransson}, C., \&
  {Soderberg}, A.~M. 2015, \apj, 810, 32

\bibitem[{{Chandra} \& {Kanekar}(2017)}]{ck17}
{Chandra}, P., \& {Kanekar}, N. 2017, \apj, 846, 111

\bibitem[{{Chen} {et~al.}(2018){Chen}, {Inserra}, {Fraser}, {Moriya}, {Schady},
  {Schweyer}, {Filippenko}, {Perley}, {Ruiter}, {Seitenzahl}, {Sollerman},
  {Taddia}, {Anderson}, {Foley}, {Jerkstrand}, {Ngeow}, {Pan}, {Pastorello},
  {Points}, {Smartt}, {Smith}, {Taubenberger}, {Wiseman}, {Young}, {Benetti},
  {Berton}, {Bufano}, {Clark}, {Della Valle}, {Galbany}, {Gal-Yam},
  {Gromadzki}, {Guti{\'e}rrez}, {Heinze}, {Kankare}, {Kilpatrick},
  {Kuncarayakti}, {Leloudas}, {Lin}, {Maguire}, {Mazzali}, {McBrien},
  {Prentice}, {Rau}, {Rest}, {Siebert}, {Stalder}, {Tonry}, \& {Yu}}]{chen+18}
{Chen}, T.~W., {Inserra}, C., {Fraser}, M., {et~al.} 2018, \apjl, 867, L31

\bibitem[{{Chevalier}(1982)}]{chevalier82a}
{Chevalier}, R.~A. 1982, \apj, 259, 302

\bibitem[{{Chevalier}(1996)}]{chevalier96}
{Chevalier}, R.~A. 1996, in Astronomical Society of the Pacific Conference
  Series, Vol.~93, Radio Emission from the Stars and the Sun, ed. A.~R.
  {Taylor} \& J.~M. {Paredes}, 125

\bibitem[{{Chevalier}(1998)}]{chevalier98}
---. 1998, \apj, 499, 810

\bibitem[{{Chevalier}(2012)}]{chevalier12}
---. 2012, \apjl, 752, L2

\bibitem[{{Chevalier} \& {Fransson}(2017)}]{cf17}
{Chevalier}, R.~A., \& {Fransson}, C. 2017, {Thermal and Non-thermal Emission
  from Circumstellar Interaction}, ed. A.~W. {Alsabti} \& P.~{Murdin}, 875

\bibitem[{{Chevalier} \& {Irwin}(2012)}]{ci12}
{Chevalier}, R.~A., \& {Irwin}, C.~M. 2012, \apjl, 747, L17

\bibitem[{{Chevalier} \& {Liang}(1989)}]{cl89}
{Chevalier}, R.~A., \& {Liang}, E.~P. 1989, \apj, 344, 332

\bibitem[{{Chugai} \& {Chevalier}(2006)}]{cc06}
{Chugai}, N.~N., \& {Chevalier}, R.~A. 2006, \apj, 641, 1051

\bibitem[{{Dwarkadas}(2005)}]{d05}
{Dwarkadas}, V.~V. 2005, \apj, 630, 892

\bibitem[{{Dwarkadas} {et~al.}(2010){Dwarkadas}, {Dewey}, \&
  {Bauer}}]{dwarkadas+10}
{Dwarkadas}, V.~V., {Dewey}, D., \& {Bauer}, F. 2010, \mnras, 407, 812

\bibitem[{{Dwarkadas} \& {Gruszko}(2012)}]{dwarkadas+12}
{Dwarkadas}, V.~V., \& {Gruszko}, J. 2012, \mnras, 419, 1515

\bibitem[{{Filippenko}(1997)}]{filippenko97}
{Filippenko}, A.~V. 1997, \araa, 35, 309

\bibitem[{{Filippenko} \& {Chornock}(2001)}]{fc01}
{Filippenko}, A.~V., \& {Chornock}, R. 2001, \iaucirc, 7737, 3

\bibitem[{{Folatelli} {et~al.}(2006){Folatelli}, {Contreras}, {Phillips},
  {Woosley}, {Blinnikov}, {Morrell}, {Suntzeff}, {Lee}, {Hamuy},
  {Gonz{\'a}lez}, {Krzeminski}, {Roth}, {Li}, {Filippenko}, {Foley},
  {Freedman}, {Madore}, {Persson}, {Murphy}, {Boissier}, {Galaz},
  {Gonz{\'a}lez}, {McCarthy}, {McWilliam}, \& {Pych}}]{folatelli+06}
{Folatelli}, G., {Contreras}, C., {Phillips}, M.~M., {et~al.} 2006, \apj, 641,
  1039

\bibitem[{{Foley} {et~al.}(2007){Foley}, {Smith}, {Ganeshalingam}, {Li},
  {Chornock}, \& {Filippenko}}]{foley07}
{Foley}, R.~J., {Smith}, N., {Ganeshalingam}, M., {et~al.} 2007, \apjl, 657,
  L105

\bibitem[{{Foreman-Mackey} {et~al.}(2013){Foreman-Mackey}, {Hogg}, {Lang}, \&
  {Goodman}}]{fm+13}
{Foreman-Mackey}, D., {Hogg}, D.~W., {Lang}, D., \& {Goodman}, J. 2013, \pasp,
  125, 306

\bibitem[{{Fruscione} {et~al.}(2006){Fruscione}, {McDowell}, {Allen},
  {Brickhouse}, {Burke}, {Davis}, {Durham}, {Elvis}, {Galle}, {Harris},
  {Huenemoerder}, {Houck}, {Ishibashi}, {Karovska}, {Nicastro}, {Noble},
  {Nowak}, {Primini}, {Siemiginowska}, {Smith}, \& {Wise}}]{fruscione+06}
{Fruscione}, A., {McDowell}, J.~C., {Allen}, G.~E., {et~al.} 2006, in Society
  of Photo-Optical Instrumentation Engineers (SPIE) Conference Series, Vol.
  6270, 62701V

\bibitem[{{Granot} \& {Ramirez-Ruiz}(2004)}]{granot+04}
{Granot}, J., \& {Ramirez-Ruiz}, E. 2004, \apjl, 609, L9

\bibitem[{{Grogin} \& {Geller}(2000)}]{grogin+00}
{Grogin}, N.~A., \& {Geller}, M.~J. 2000, \aj, 119, 32

\bibitem[{{Harris} {et~al.}(2016){Harris}, {Nugent}, \& {Kasen}}]{harris+16}
{Harris}, C.~E., {Nugent}, P.~E., \& {Kasen}, D.~N. 2016, \apj, 823, 100

\bibitem[{{Hatchett} {et~al.}(1976){Hatchett}, {Buff}, \&
  {McCray}}]{hatchett+76}
{Hatchett}, S., {Buff}, J., \& {McCray}, R. 1976, \apj, 206, 847

\bibitem[{{Houck}(2005)}]{houck+05}
{Houck}, J.~C. 2005, Astronomical Society of the Pacific Conference Series,
  Vol. 332, {X-Ray Emission from SN 1986J}, ed. R.~{Humphreys} \& K.~{Stanek},
  429

\bibitem[{{Immler} \& {Kuntz}(2005)}]{sk+05}
{Immler}, S., \& {Kuntz}, K.~D. 2005, \apjl, 632, L99

\bibitem[{{Inserra} {et~al.}(2014){Inserra}, {Smartt}, {Scalzo}, {Fraser},
  {Pastorello}, {Childress}, {Pignata}, {Jerkstrand }, {Kotak}, {Benetti},
  {Della Valle}, {Gal-Yam}, {Mazzali}, {Smith}, {Sullivan}, {Valenti}, {Yaron},
  {Young}, \& {Reichart}}]{inserra+14}
{Inserra}, C., {Smartt}, S.~J., {Scalzo}, R., {et~al.} 2014, \mnras, 437, L51

\bibitem[{{Kale} {et~al.}(2018){Kale}, {Parekh}, \& {Dwarakanath}}]{ruta}
{Kale}, R., {Parekh}, V., \& {Dwarakanath}, K.~S. 2018, \mnras, 480, 5352

\bibitem[{{Karamehmetoglu} {et~al.}(2019){Karamehmetoglu}, {Fransson},
  {Sollerman}, {Tartaglia}, {Taddia}, {De}, {Fremling}, {Bagdasaryan},
  {Barbarino}, {Bellm}, {Dekaney}, {Dugas}, {Giomi}, {Goobar}, {Graham}, {Ho},
  {Laher}, {Masci}, {Neill}, {Perley}, {Riddle}, {Rusholme}, \&
  {Soumagnac}}]{Karamehmetoglu19}
{Karamehmetoglu}, E., {Fransson}, C., {Sollerman}, J., {et~al.} 2019, arXiv
  e-prints, arXiv:1910.06016

\bibitem[{{Margutti} {et~al.}(2014){Margutti}, {Milisavljevic}, {Soderberg},
  {Chornock}, {Zauderer}, {Murase}, {Guidorzi}, {Sanders}, {Kuin}, {Fransson},
  {Levesque}, {Chandra}, {Berger}, {Bianco}, {Brown}, {Challis},
  {Chatzopoulos}, {Cheung}, {Choi}, {Chomiuk}, {Chugai}, {Contreras}, {Drout},
  {Fesen}, {Foley}, {Fong}, {Friedman}, {Gall}, {Gehrels}, {Hjorth}, {Hsiao},
  {Kirshner}, {Im}, {Leloudas}, {Lunnan}, {Marion}, {Martin}, {Morrell},
  {Neugent}, {Omodei}, {Phillips}, {Rest}, {Silverman}, {Strader},
  {Stritzinger}, {Szalai}, {Utterback}, {Vinko}, {Wheeler}, {Arnett},
  {Campana}, {Chevalier}, {Ginsburg}, {Kamble}, {Roming}, {Pritchard}, \&
  {Stringfellow}}]{margutti+14}
{Margutti}, R., {Milisavljevic}, D., {Soderberg}, A.~M., {et~al.} 2014, \apj,
  780, 21

\bibitem[{{Margutti} {et~al.}(2017){Margutti}, {Kamble}, {Milisavljevic},
  {Zapartas}, {de Mink}, {Drout}, {Chornock}, {Risaliti}, {Zauderer},
  {Bietenholz}, {Cantiello}, {Chakraborti}, {Chomiuk}, {Fong}, {Grefenstette},
  {Guidorzi}, {Kirshner}, {Parrent}, {Patnaude}, {Soderberg}, {Gehrels}, \&
  {Harrison}}]{margutti+17}
{Margutti}, R., {Kamble}, A., {Milisavljevic}, D., {et~al.} 2017, \apj, 835,
  140

\bibitem[{{Mauerhan} {et~al.}(2018{\natexlab{a}}){Mauerhan}, {Filippenko},
  {Zheng}, {Brink}, {Graham}, {Shivvers}, \& {Clubb}}]{mauerhan18}
{Mauerhan}, J.~C., {Filippenko}, A.~V., {Zheng}, W., {et~al.}
  2018{\natexlab{a}}, \mnras, 478, 5050

\bibitem[{{Mauerhan} {et~al.}(2018{\natexlab{b}}){Mauerhan}, {Filippenko},
  {Zheng}, {Brink}, {Graham}, {Shivvers}, \& {Clubb}}]{mauerhan+18}
---. 2018{\natexlab{b}}, \mnras, 478, 5050

\bibitem[{{McMullin} {et~al.}(2007){McMullin}, {Waters}, {Schiebel}, {Young},
  \& {Golap}}]{casa}
{McMullin}, J.~P., {Waters}, B., {Schiebel}, D., {Young}, W., \& {Golap}, K.
  2007, Astronomical Society of the Pacific Conference Series, Vol. 376, {CASA
  Architecture and Applications}, ed. R.~A. {Shaw}, F.~{Hill}, \& D.~J. {Bell},
  127

\bibitem[{{Milisavljevic} {et~al.}(2015){Milisavljevic}, {Margutti}, {Kamble},
  {Patnaude}, {Raymond}, {Eldridge}, {Fong}, {Bietenholz}, {Challis},
  {Chornock}, {Drout}, {Fransson}, {Fesen}, {Grindlay}, {Kirshner}, {Lunnan},
  {Mackey}, {Miller}, {Parrent}, {Sand ers}, {Soderberg}, \&
  {Zauderer}}]{milisavljevic+15}
{Milisavljevic}, D., {Margutti}, R., {Kamble}, A., {et~al.} 2015, \apj, 815,
  120

\bibitem[{{Modjaz} {et~al.}(2016){Modjaz}, {Liu}, {Bianco}, \&
  {Graur}}]{modjaz+16}
{Modjaz}, M., {Liu}, Y.~Q., {Bianco}, F.~B., \& {Graur}, O. 2016, \apj, 832,
  108

\bibitem[{{Ofek} {et~al.}(2014){Ofek}, {Zoglauer}, {Boggs}, {Barri{\'e}re},
  {Reynolds}, {Fryer}, {Harrison}, {Cenko}, {Kulkarni}, {Gal-Yam}, {Arcavi},
  {Bellm}, {Bloom}, {Christensen}, {Craig}, {Even}, {Filippenko},
  {Grefenstette}, {Hailey}, {Laher}, {Madsen}, {Nakar}, {Nugent}, {Stern},
  {Sullivan}, {Surace}, \& {Zhang}}]{ofek+14}
{Ofek}, E.~O., {Zoglauer}, A., {Boggs}, S.~E., {et~al.} 2014, \apj, 781, 42

\bibitem[{{Papenkova} {et~al.}(2001){Papenkova}, {Li}, {Wray}, {Chleborad}, \&
  {Schwartz}}]{papenkova+01}
{Papenkova}, M., {Li}, W.~D., {Wray}, J., {Chleborad}, C.~W., \& {Schwartz}, M.
  2001, \iaucirc, 7722

\bibitem[{{Paragi} {et~al.}(2005){Paragi}, {Garrett}, {Paczy{\'n}ski},
  {Kouveliotou}, {Szomoru}, {Reynolds}, {Parsley}, \& {Ghosh}}]{paragi+05}
{Paragi}, Z., {Garrett}, M.~A., {Paczy{\'n}ski}, B., {et~al.} 2005, \memsai,
  76, 570

\bibitem[{{Pastorello} {et~al.}(2007{\natexlab{a}}){Pastorello}, {Smartt},
  {Mattila}, {Eldridge}, {Young}, {Itagaki}, {Yamaoka}, {Navasardyan},
  {Valenti}, {Patat}, {Agnoletto}, {Augusteijn}, {Benetti}, {Cappellaro},
  {Boles}, {Bonnet-Bidaud}, {Botticella}, {Bufano}, {Cao}, {Deng}, {Dennefeld},
  {Elias-Rosa}, {Harutyunyan}, {Keenan}, {Iijima}, {Lorenzi}, {Mazzali},
  {Meng}, {Nakano}, {Nielsen}, {Smoker}, {Stanishev}, {Turatto}, {Xu}, \&
  {Zampieri}}]{pastorello+07}
{Pastorello}, A., {Smartt}, S.~J., {Mattila}, S., {et~al.} 2007{\natexlab{a}},
  \nat, 447, 829

\bibitem[{{Pastorello} {et~al.}(2007{\natexlab{b}}){Pastorello}, {Smartt},
  {Mattila}, {Eldridge}, {Young}, {Itagaki}, {Yamaoka}, {Navasardyan},
  {Valenti}, {Patat}, {Agnoletto}, {Augusteijn}, {Benetti}, {Cappellaro},
  {Boles}, {Bonnet-Bidaud}, {Botticella}, {Bufano}, {Cao}, {Deng}, {Dennefeld},
  {Elias-Rosa}, {Harutyunyan}, {Keenan}, {Iijima}, {Lorenzi}, {Mazzali},
  {Meng}, {Nakano}, {Nielsen}, {Smoker}, {Stanishev}, {Turatto}, {Xu}, \&
  {Zampieri}}]{pastorello07}
---. 2007{\natexlab{b}}, \nat, 447, 829

\bibitem[{{Pooley} \& {Lewin}(2004)}]{pooley+04}
{Pooley}, D., \& {Lewin}, W.~H.~G. 2004, \iaucirc, 8323

\bibitem[{{Pooley} {et~al.}(2019){Pooley}, {Wheeler}, {Vink{\'o}}, {Dwarkadas},
  {Szalai}, {Silverman}, {Griesel}, {McCullough}, {Marion}, \&
  {MacQueen}}]{pooley+19}
{Pooley}, D., {Wheeler}, J.~C., {Vink{\'o}}, J., {et~al.} 2019, \apj, 883, 120

\bibitem[{{Prasad} \& {Chengalur}(2012)}]{prasad12}
{Prasad}, J., \& {Chengalur}, J. 2012, Experimental Astronomy, 33, 157

\bibitem[{{Puls} {et~al.}(2008){Puls}, {Vink}, \& {Najarro}}]{puls+08}
{Puls}, J., {Vink}, J.~S., \& {Najarro}, F. 2008, \aapr, 16, 209

\bibitem[{{Rupen} {et~al.}(1987){Rupen}, {van Gorkom}, {Knapp}, {Gunn}, \&
  {Schneider}}]{rupen+87}
{Rupen}, M.~P., {van Gorkom}, J.~H., {Knapp}, G.~R., {Gunn}, J.~E., \&
  {Schneider}, D.~P. 1987, \aj, 94, 61

\bibitem[{{Schinzel} {et~al.}(2008){Schinzel}, {Taylor}, {Stockdale}, {Granot},
  \& {Ramirez-Ruiz}}]{schinzel+08}
{Schinzel}, F., {Taylor}, G.~B., {Stockdale}, C.~J., {Granot}, J., \&
  {Ramirez-Ruiz}, E. 2008, in The role of VLBI in the Golden Age for Radio
  Astronomy, 94

\bibitem[{{Schlegel} {et~al.}(1998){Schlegel}, {Finkbeiner}, \&
  {Davis}}]{schlegel+98}
{Schlegel}, D.~J., {Finkbeiner}, D.~P., \& {Davis}, M. 1998, \apj, 500, 525

\bibitem[{{Schlegel} \& {Petre}(2006)}]{schlegel+06}
{Schlegel}, E.~M., \& {Petre}, R. 2006, \apj, 646, 378

\bibitem[{{Schr{\o}der} {et~al.}(2020){Schr{\o}der}, {MacLeod}, {Loeb},
  {Vigna-G{\'o}mez}, \& {Mandel}}]{sophie+20}
{Schr{\o}der}, S.~L., {MacLeod}, M., {Loeb}, A., {Vigna-G{\'o}mez}, A., \&
  {Mandel}, I. 2020, \apj, 892, 13

\bibitem[{{Shiode} \& {Quataert}(2014)}]{sq14}
{Shiode}, J.~H., \& {Quataert}, E. 2014, \apj, 780, 96

\bibitem[{{Shivvers} {et~al.}(2019){Shivvers}, {Filippenko}, {Silverman},
  {Zheng}, {Foley}, {Chornock}, {Barth}, {Cenko}, {Clubb}, {Fox},
  {Ganeshalingam}, {Graham}, {Kelly}, {Kleiser}, {Leonard}, {Li}, {Matheson},
  {Mauerhan}, {Modjaz}, {Serduke}, {Shields}, {Steele}, {Swift}, {Wong}, \&
  {Yuk}}]{shivvers19}
{Shivvers}, I., {Filippenko}, A.~V., {Silverman}, J.~M., {et~al.} 2019, \mnras,
  482, 1545

\bibitem[{{Smith}(2014)}]{smith14}
{Smith}, N. 2014, \araa, 52, 487

\bibitem[{{Smith}(2017)}]{smith17}
---. 2017, {Interacting Supernovae: Types IIn and Ibn}, ed. A.~W. {Alsabti} \&
  P.~{Murdin}, 403

\bibitem[{{Smith} \& {Owocki}(2006)}]{so06}
{Smith}, N., \& {Owocki}, S.~P. 2006, \apjl, 645, L45

\bibitem[{{Smith} {et~al.}(2018){Smith}, {Andrews}, {Rest}, {Bianco}, {Prieto},
  {Matheson}, {James}, {Smith}, {Strampelli}, \& {Zenteno}}]{smith+18}
{Smith}, N., {Andrews}, J.~E., {Rest}, A., {et~al.} 2018, \mnras, 480, 1466

\bibitem[{{Soderberg} {et~al.}(2004){Soderberg}, {Gal-Yam}, \&
  {Kulkarni}}]{soderberg+04}
{Soderberg}, A.~M., {Gal-Yam}, A., \& {Kulkarni}, S.~R. 2004, GRB Coordinates
  Network, 2586

\bibitem[{{Spitzer} \& {Arny}(1978)}]{spitzer78}
{Spitzer}, L., \& {Arny}, T.~T. 1978, American Journal of Physics, 46, 1201

\bibitem[{{Stockdale} {et~al.}(2004){Stockdale}, {Van Dyk}, {Sramek}, {Weiler},
  {Panagia}, {Rupen}, \& {Paczynski}}]{stockdale+04}
{Stockdale}, C.~J., {Van Dyk}, S.~D., {Sramek}, R.~A., {et~al.} 2004, \iaucirc,
  8282

\bibitem[{{Stockdale} {et~al.}(2005){Stockdale}, {Kaster}, {Sjouwerman},
  {Rupen}, {Marti-Vidal}, {Marcaide}, {van Dyk}, {Weiler}, {Paczynski}, \&
  {Panagia}}]{stockdale+05}
{Stockdale}, C.~J., {Kaster}, B., {Sjouwerman}, L.~O., {et~al.} 2005, \iaucirc,
  8472

\bibitem[{{van Dyk}(2010)}]{vandyk10}
{van Dyk}, S.~D. 2010, in Astronomical Society of the Pacific Conference
  Series, Vol. 425, Hot and Cool: Bridging Gaps in Massive Star Evolution, ed.
  C.~{Leitherer}, P.~D. {Bennett}, P.~W. {Morris}, \& J.~T. {Van Loon}, 73

\bibitem[{{van Dyk} {et~al.}(1993){van Dyk}, {Weiler}, {Sramek}, \&
  {Panagia}}]{vandyk+93}
{van Dyk}, S.~D., {Weiler}, K.~W., {Sramek}, R.~A., \& {Panagia}, N. 1993,
  \apjl, 419, L69

\bibitem[{{van Marle} {et~al.}(2010){van Marle}, {Smith}, {Owocki}, \& {van
  Veelen}}]{vanmarle+10}
{van Marle}, A.~J., {Smith}, N., {Owocki}, S.~P., \& {van Veelen}, B. 2010,
  \mnras, 407, 2305

\bibitem[{{Vinko} {et~al.}(2017){Vinko}, {Pooley}, {Silverman}, {Wheeler},
  {Szalai}, {Kelly}, {MacQueen}, {Marion}, \& {S{\'a}rneczky}}]{vinko+17}
{Vinko}, J., {Pooley}, D., {Silverman}, J.~M., {et~al.} 2017, \apj, 837, 62

\bibitem[{{Weiler} {et~al.}(2002){Weiler}, {Panagia}, {Montes}, \&
  {Sramek}}]{weiler+02}
{Weiler}, K.~W., {Panagia}, N., {Montes}, M.~J., \& {Sramek}, R.~A. 2002,
  \araa, 40, 387

\bibitem[{{Weiler} {et~al.}(1990){Weiler}, {Panagia}, \& {Sramek}}]{weiler+90}
{Weiler}, K.~W., {Panagia}, N., \& {Sramek}, R.~A. 1990, \apj, 364, 611

\bibitem[{{Wellons} {et~al.}(2012){Wellons}, {Soderberg}, \&
  {Chevalier}}]{wellons+12}
{Wellons}, S., {Soderberg}, A.~M., \& {Chevalier}, R.~A. 2012, \apj, 752, 17

\bibitem[{{Williams} {et~al.}(2002){Williams}, {Panagia}, {Van Dyk}, {Lacey},
  {Weiler}, \& {Sramek}}]{williams+02}
{Williams}, C.~L., {Panagia}, N., {Van Dyk}, S.~D., {et~al.} 2002, \apj, 581,
  396

\end{thebibliography}
\bibliographystyle{apj}

\end{document}